\documentclass[aps,pre,twocolumn,eqsecnum,showpacs]{revtex4}
\usepackage{amsmath}
\usepackage{bm}
\usepackage{rev2002a}
\usepackage{macro1604o}
\newcommand{\Vmax}{{V_{\text{max}}}}
\newcommand{\muR}{{\mu_{\text{r}}}}
\usepackage{graphicx}
\graphicspath{{Fig.dummy/}{newfig/}{../lin2d15v4/}{.}}
\usepackage{CJK}
\newcommand{\IncFig}[2][clip]{\includegraphics[#1]{#2}}
\newcommand{\SubFig}[2][5.5cm]{\raisebox{#1}{(#2)}\hfill}

\newcommand{\Da}{D_\alpha}
\newcommand{\rInit}{\mb{r}^{\mathrm{init}}}
\newcommand{\xInit}{x^{\mathrm{init}}}
\newcommand{\abs}[1]{\left|{#1}\right|}
\newcommand{\Order}{\mathcal{O}}
\newcommand{\sumPQK}{\sum_{\mb{p}+\mb{q}+\mb{k}=\mb{0}}}
\newcommand{\Vx}[6]{\Vertex_{{#1},\mb{#2}}^{#3#4}(\mb{#5},\mb{#6})}
\newcommand{\V}[6]{\Vx{\mathrm{#1}}{#2}{\mathrm{#3}}{\mathrm{#4}}#5#6}
\newcommand{\fd}{\check{f}_\text{d}}
\newcommand{\fr}{\check{f}_\text{r}}
\newcommand{\ek}{\mb{e}_{\mb{k}}}
\begin{document}
\begin{CJK*}{JIS}{}  
\title{Calculation of displacement correlation tensor 
       indicating vortical cooperative motion 
       in two-dimensional colloidal liquids}
\author{\surname{Ooshida} Takeshi}
\email[E-mail:~]{ooshida@damp.tottori-u.ac.jp}
\affiliation{%
   Department of Mechanical and Aerospace Engineering,
   Tottori University, Tottori 680-8552, Japan}
\author{Susumu \surname{Goto}}
\affiliation{%
   Graduate School of Engineering Science,  
   Osaka University, 
   \relax{Toyonaka, Osaka 560-8531, Japan}}
\author{Takeshi \surname{Matsumoto}}
\affiliation{%
   Division of Physics and Astronomy, Graduate School of Science, 
   Kyoto University, 
   \relax{Kyoto 606-8502, Japan}}
\author{Michio \surname{Otsuki}}
\affiliation{%
   Department of Materials Science, 
   Shimane University, 
   \relax{Matsue 690-8504, Japan}}



\begin{abstract}
As an indicator of cooperative motion 
  in a system of Brownian particles 
  that models two-dimensional colloidal liquids, 
  displacement correlation tensor is calculated analytically 
  and compared with numerical results.
The key idea for the analytical calculation 
  is to relate the displacement correlation tensor, 
  which is a kind of four-point space--time correlation,  
  to the Lagrangian two-time correlation 
  of the deformation gradient tensor. 
Tensorial treatment of the statistical quantities, 
  including the displacement correlation itself,  
  allows capturing the vortical structure 
  of the cooperative motion. 
The calculated displacement correlation 
  also implies a negative longtime tail 
  in the velocity autocorrelation,  
  which is a manifestation of the cage effect.
Both the longitudinal and transverse components 
  of the displacement correlation 
  are found to be expressible in terms of a similarity variable, 
  suggesting that the cages are nested 
  to form a self-similar structure in the space--time.
\end{abstract}

\maketitle
\end{CJK*}

\section{Introduction}
\label{sec:intro}

Cooperative motions of densely packed particles
  \cite{Hiwatari.JNCS235,Yamamoto.PRE58}
  are now generally recognized 
  as an essential ingredient
  of the statistical mechanics 
  of such a system,
  and their characterization and quantification 
  have comprised one of the central problems in the area
  \cite{Berthier.RMP83,Berthier.Book2011}.
These cooperative motions 
  are often found to consist of swirls or vortices
  \cite{Doliwa.PRE61,Brito.JCP131,Sota.JKPS54,Ghosh.PRL104}, 
  reminiscent of the vortices of momentum 
  observed in other systems \cite{Alder.PRA1,van-Noije.PRE61}. 
Similar vortex patterns 
  are now known to appear frequently 
  in systems of self-propelled units (both living and nonliving)
  \cite{Vicsek.PhysRep517}. 
Having observed these conspicuous vortices in systems 
  of molecular, colloidal, granular and self-propelled particles, 
  we raise the following questions: 
Can we 
  incorporate such vortices 
  into the statistical theory of the particles, 
  so as to understand certain aspects of their dynamics
  \emph{quantitatively}?
In particular, 
  is it possible to calculate statistical quantities 
  indicating vortical cooperative motion 
  in systems in which momentum is \emph{not} conserved, 
  such as colloidal liquids subject to Langevin dynamics?

The question of quantitative theory with vortices 
  is quite challenging, 
  as is seen in analogous attempts 
  in the research of fluid turbulence 
  that is filled with vortex tubes, blobs and sheets
  \cite{Frisch.Book1995}.
The velocity fluctuations of high Reynolds-number turbulence 
  are characterized by statistical scaling laws
  \cite{Frisch.Book1995,Batchelor.Book1953,Hinze.Book1975,%
  Landau.fluid}, 
  which are not yet understood 
  on the basis of the first principles, 
  namely by solving the Navier--Stokes equation.
As an approach 
  based on vortical solutions to the Navier--Stokes equation, 
  we may mention 
  theoretical calculation of velocity statistics 
  from superposition of a specific vortical pattern 
  \cite{Saffman.PhF8}.
It is criticizable, however,
  that the choice of the vortex model 
  and the assumption on the distribution of the vortex patterns 
  are rather arbitrary,
  as was pointed out by the authors themselves 
  \cite{Pullin.ARFM30}. 


In the background of the second question, 
  motivated by the suspect 
  that momentum conservation is not essential 
  to the vortical cooperative motion,
  there is a fact 
  that different space--time correlations 
  have been devised for different systems. 
While the equal-time correlation of velocity
  is already informative in some systems, 
  a more elaborate statistical quantity 
  is often needed for other systems.
In the case of colloidal suspensions
  as model glassy systems,
  the most widely used idea 
  is to focus on the density field, $\rho(\mb{r},t)$,  
  and define its four-point space--time correlation  
  by analogy with spin glasses
  \cite{Dasgupta.EPL15,Glotzer.JCP112,Berthier.RMP83}. 
In this context, 
  the case of dense liquids 
  subject to Newtonian molecular dynamics 
  deserves special consideration.
Indeed, momentum is conserved in such systems,
  but their longtime behavior 
  is known to be basically the same 
  as that of colloidal glasses 
  \cite{Szamel.PRA44,Gleim.PRL81,Berthier.JCP126}.
The cooperative motions in dense molecular liquids 
  seem rather to stem from mass conservation, 
  and a popular approach to this cooperativity 
  is, again, based on four-point space--time correlations 
  of $\rho(\mb{r},t)$.

A disadvantage of this approach is that, 
  since $\rho(\mb{r},t)$ is scalar, 
  it lacks direct access to 
  the directional aspect of the cooperative motion.
Admittedly, 
  some refined scalar descriptions
  such as many-particle density approach \cite{Szamel.JCP127} 
  are possible, 
  but they are difficult to illustrate pictorially 
  and therefore not very intuitive.

Another approach to the cooperative motion 
  is based on the displacement of the particles.
Let us denote the displacement of the $i$-th particle
  for the time interval from $s$ to $t$
  with 
\begin{equation}
  \mb{R}_i = \mb{R}_i(t,s) = \mb{r}_i(t) - \mb{r}_i(s)
  \label{R=},
\end{equation}  
  where $\mb{r}_i$ is the position vector of the particle.
Indeed, $\mb{R}_i$ includes 
  information of the direction of motion, 
  but again this information was discarded  
  in most of the existing studies, 
  in which statistics of scalar quantities 
  such as $\Av{{\mb{R}_i}\cdot{\mb{R}_j}}$ 
  \cite{Muranaka.PRE51,Hiwatari.JNCS235}
  or $\Av{\abs{\mb{R}_i}\abs{\mb{R}_j}}$   
  \cite{Donati.PRL82}
  were computed.

The directionality, however, 
  is crucial for understanding the cooperative motion,
  which often consists of vortices
  \cite{Brito.JCP131,Sota.JKPS54,Ghosh.PRL104}
  as we have already noticed.
Conversely, 
  the space--time structure of the vortical motion 
  is expected to contain 
  information of the directional statistics of $\mb{R}_i$.  
  In the context of glassy liquids,
  recently there appeared several studies 
  on such statistics 
  in connection with solid-like elasticity 
  \cite{Klix.PRL109,Flenner.PRL114,Ikeda.PRE92,Bernini.JCP144}.
Now let us focus on a more pictorial illustration
  of directional statistics 
  given 
  by Doliwa and Heuer \cite{Doliwa.PRE61}: 
  in their Fig.~8,  
  the particles ``in front'' and ``in back'' of the central one, 
  i.e.\ 
  the particles with the relative position vector 
  parallel to $\mb{R}_i$, 
  have a very long-ranged directional correlation, 
  while the ``sideways'' particles 
  form a kind of back\-flow structure with a pair of swirls
  (as we will see later in Fig.~\ref{Fig:RR.2D}). 
In other words, 
  with $\mb{r}_{ij} = \mb{r}_j -\mb{r}_i$
  denoting the relative position vector,
  the displacement correlation 
  for $\mb{r}_{ij}$ parallel to $\mb{R}_i$
  is considerably different 
  from that for $\mb{r}_{ij}$ 
  perpendicular to $\mb{R}_i$.

Mathematically, 
  there is a simple way to retain 
  the directional information in $\mb{R}_i$: 
  instead of the inner product 
  in $\Av{{\mb{R}_i}\cdot{\mb{R}_j}}$ 
  in the preceding studies
  \cite{Muranaka.PRE51,Hiwatari.JNCS235},
  we may adopt the tensorial product 
  and study the \emph{displacement correlation tensor}
  \cite{Ooshida.PRE88,Ooshida.BRL11}, 
  which we denote symbolically 
  with $\Av{\mb{R}_i\otimes\mb{R}_j}$ 
  or   $\Av{\mb{R}  \otimes\mb{R}  }$
  (a more precise definition will be given later). 
The displacement correlation tensor, on one hand,
  includes essentially the same information 
  as the parallel and perpendicular correlations 
  defined by Doliwa and Heuer \cite{Doliwa.PRE61}. 
On the other hand, 
  as the definition of $\Av{\mb{R}\otimes\mb{R}}$ 
  is free from conditional angular summation 
  used by Doliwa and Heuer \cite{Doliwa.PRE61},
  it is more amenable to analytical calculation 
  \cite{Ooshida.BRL11}. 

In the present paper, 
  we compare 
  the analytical calculation of $\Av{\mb{R}\otimes\mb{R}}$ 
  \cite{Ooshida.BRL11}
  with the numerical solutions of the Langevin dynamics 
  as a model of two-dimensional (2D) colloidal liquids. 
The paper is organized 
  as follows: 
In Sec.~\ref{sec:setup},  
  the governing equation is specified 
  together with simulation details.
A continuum version of the governing equation 
  is also given 
  and some statistical quantities are defined, 
  which forms a basis for the analytical calculation.
Subsequently, 
  in Sec.~\ref{sec:DC}, 
  the displacement correlation tensor 
  is defined and computed numerically.
Analytical calculation
  is then explained in Sec.~\ref{sec:theory}, 
  which is based on a rather crude approximation 
  and intended as a springboard 
  for a more elaborate theory in the future.
Comparison between the numerical and analytical results
  is made in Sec.~\ref{sec:compare}; 
  considering the limitations 
  of the approximation,  
  we find the two results to agree to a surprising extent, 
  except for the short-range behavior.
Possible relevance to glassy dynamics and future directions 
  are discussed 
  in Sec.~\ref{sec:discuss},
  and finally, Sec.~\ref{sec:conc}
  is allotted for concluding remarks.

\section{Setup}
\label{sec:setup}

\subsection{Specification of the particle system}

We consider a system 
  consisting of $N$ Brownian particles 
  in an $\nd$-dimensional periodic box of the size $L^\nd$,
  focusing on the 2D case
  and also paying some attention 
  to the one-dimensional (1D) case,
  so that $\nd = 2$ or $\nd = 1$.
The system
  is statistically steady and homogeneous, 
  with the mean density $\rho_0 = N/L^\nd$,
  being in equilibrium at the temperature $T$.
The position vectors of the particles,
  denoted by $\mb{r}_i$ with $i = 1, 2, \ldots, N$, 
  are governed 
  by the over\-damped Langevin equation  
\begin{equation}
  \mu \dot{\mb{r}}_i  
   = - \dd{}{\mb{r}_i} \sum_{j<k} V_{jk}
     + \mu \mb{f}_i(t)
  \label{Langevin-.r},
\end{equation}
  where 
  $\mu$ is the drag coefficient, 
  and 
  $V_{jk}$ is the interaction potential 
  between the $j$-th and $k$-th particles.
The random forcing is Gaussian
  with the variance 
\begin{equation}
  \Av{\mb{f}_i(t) \otimes \mb{f}_j(t')}
  = 2D \delta_{ij} \delta(t-t') \openone
  \label{fvect}
\end{equation}%
  where $D = \kT/\mu$.
The pair potential $V_{jk}$
  is repulsive and short-ranged,
  so that the particle diameter can be defined 
  through the excluded volume effect.
For concreteness, 
  as a function of $r_{jk} = \abs{\mb{r}_k - \mb{r}_j}$,
  we specify the potential 
  as \cite{Ikeda.PRE92,Berthier.EPL86,note.V}
\begin{equation}  
  V_{jk} =
  \begin{cases}
    \Vmax\left(1-\dfrac{|r_{jk}|}{\sigma}\right)^2 
    \relax    & (|r_{jk}| \le\sigma) \\
    0         & (|r_{jk}|  > \sigma) 
  \end{cases}  
  \label{V=}
\end{equation}  
  with $\sigma$ 
  denoting the diameter of the particles.
To make $V_{jk}$ 
  close to the hardcore potential, 
  sufficiently large values of $\Vmax$ were used:
  the value 
  is specified as $\Vmax/\kT = 50$,
  unless noted otherwise.

Aside from the density $\rho_0 = N/L^2$ for $\nd=2$,  
  we denote the area fraction 
  with $\phi 
  = (\pi/4)\sigma^2 N/L^2 = (\pi/4)\sigma^2 \rho_0$.  
Numerical calculations are performed 
  with $\phi = 0.50$ and $N = 4000$, 
  unless specified otherwise.
The initial condition is prepared 
  by starting from a random configuration 
  and equilibrating the system 
  under the Langevin dynamics 
  with the target temperature $T$,  
  for sufficiently long waiting time 
  (typically $20\,\sigma^2/D$).
The computed results are nondimensionalized 
  so that $\sigma$, $D$, and $\mu$ become equal to unity.

\subsection{Continuum description}

For the theoretical approach in Sec.~\ref{sec:theory},
  the overdamped Langevin equation for $\mb{r}_i(t)$ 
  is rewritten in terms of the density field, 
\begin{equation}
  \rho = \rho(\mb{r},t) = \sum_j \rho_j(\mb{r},t)
  \label{rho=},
\end{equation}
  where $\rho_j(\mb{r},t) 
  = \delta^\nd(\mb{r}-\mb{r}_j(t))$
  at the finer level of description 
  as the operator acting on $\mb{r}_j(t)$ 
  \cite{Chaikin.Book1995},
  with $\delta^\nd(~\cdot~)$ denoting 
  the $\nd$-dimensional delta function,
  but in practice $\rho_j(\mb{r},t)$ should be regarded 
  as its average over the local equilibrium ensemble
  \cite{Das.PRE88,Bidhoodi.PRE92.DK}. 
The equation for $\rho$
  is customarily referred to 
  as the Dean--Kawasaki equation
  \cite{Dean.JPhAMG29,Kawasaki.PhysicaA208,%
  Kawasaki.JStatPhys93,Das.Book2011,Kim.PRE89};
  it is a kind of stochastic diffusion equation
  \cite{Gardiner.Book2009,Mikhailov.Book1996,van-Kampen.Book2007},  
  analogous to nonlinear stochastic equations 
  describing mesoscopic kinetics of phase separation 
  \cite{Kawasaki.Synergetics1973,Kawasaki.PhysicaA201,Hildebrand.JPC49},
  though the appellation ``Dean--Kawasaki'' 
  is limited to the one with a specific form of nonlinearity 
  prescribed below.
It is convenient to express it   
  as a set of two equations,
  consisting of the continuity equation
\begin{equation}
  \dt\rho + \nabla\cdot\mb{Q} = 0  \label{+cont}
\end{equation}
  and a stochastic equation for the flux, 
\begin{equation}
  \mb{Q} 
  = -D\,\left( \nabla\rho + \frac{\rho}{\kT}\nabla{U} \right)
  + \sum_j \rho_j(\mb{r},t) \mb{f}_j(t) 
  \label{DK.Q},
\end{equation}
  where $\mb{f}_j(t)$ denotes the random forcing 
  subject to Eq.~(\ref{fvect})
  and
\begin{equation}
  U = U[\rho](\mb{r}) 
  = \int \Veff(r_*)\rho(\tilde{\mb{r}})\D^2\tilde{\mb{r}} 
  \quad (\text{for}~\nd=2)
  \label{U=} 
\end{equation}
  describes the interaction of the particles, 
  with $\Veff$ denoting the effective potential 
  based on the direct correlation function 
  \cite{Dhont.Book1996,Das.Book2011,Das.RMP76}, 
  related to the static structure factor $S(k)$ 
  explained below [in Eq.~(\ref{S=})],
  and $r_* = \left|{\mb{r} - \tilde{\mb{r}}}\right|$.
Note that $\Veff$ appears 
  as a result of the coarse-graining 
  \cite{Das.PRE88,Bidhoodi.PRE92.DK}.

Elimination of $\mb{Q}$ 
  from Eqs.~(\ref{+cont}) and (\ref{DK.Q}) 
  yields a single equation for $\rho(\mb{r},t)$
  which is often taken as a starting point 
  for nonlinear theory of glassy dynamics
  \cite{Miyazaki.JPA38,Andreanov.JStat2006,%
  Kim.JPhA40,Kim.JStat2008,Kim.PRE89}.
In Sec.~\ref{sec:theory}, however,
  we will prefer to start from Eq.~(\ref{DK.Q})
  without this elimination,
  in the hope that the presence of $\mb{Q}$ will be helpful 
  in capturing the directional aspect of the cooperative motion.
We emphasize 
  that $\mb{Q}$ enters the theory 
  not because it is momentum 
  which may or may not be conserved, 
  but because it is the flux 
  of the conserved quantity $\rho$.

The static structure factor
  is defined by 
\begin{equation}
   S(k) = \frac1N\sum_{i,j}
   \Av{ e^{\II\mb{k}\cdot\left(\mb{r}_j-\mb{r}_i\right)}}
   \quad (\mb{k}\ne\mb{0})
   \label{S=}  
\end{equation}
  based on a snapshot of the particle configuration.
Evidently,  
  $S(k)$ equals
  the initial value 
  of the intermediate scattering function
  (the dynamical structure factor),
\begin{equation}
   F(k,t) = \frac1N\sum_{i,j}
   \Av{ e^{\II\mb{k}\cdot\left[\mb{r}_j(t)-\mb{r}_i(0)\right]}}
   \quad (\mb{k}\ne\mb{0})
   \label{F=}.  
\end{equation}  
For colloidal systems modeled by Eq.~(\ref{Langevin-.r})
  with constant scalar $\mu$ (i.e.\ 
  without hydrodynamic interaction),
  the short-time behavior of $F(k,t)$
  is known to be described by 
\begin{equation}
  F(k,t) = S(k) e^{-\Dc\mb{k}^2 t}, \quad
  \Dc = \Dc(k) = \frac{D}{S(k)}
  \label{Dc}, 
\end{equation}
  with $\Dc$ referred to 
  as the (short-time) collective diffusion coefficient 
  \cite{Dhont.Book1996}.
To be consistent with Eq.~(\ref{Dc}), 
  the coefficient of the linear term 
  of the Dean--Kawasaki equation in the Fourier representation 
  must be $-{\Dc}\mb{k}^2$. 

For later convenience, we define 
\begin{equation}
  \ell_0 = \frac{L}{N^{1/\nd}}
  = 
  \begin{cases}
    1/\rho_0        & (\nd=1)  \\
    1/\sqrt{\rho_0} & (\nd=2)
  \end{cases}
  \label{len0}
\end{equation}
  which is on the order of the typical distance 
  between the centers of the neighboring particles.

\section{Displacement correlation tensor}
\label{sec:DC}

For the system of colloidal particles 
  specified in the previous section
  and governed by Eq.~(\ref{Langevin-.r}), 
  now let us devise a statistical quantity  
  which is at once indicative of the cooperative motion 
  and amenable to analytical calculation.
The quantity of our choice for it 
  is the displacement correlation tensor.

\subsection{Definition}
\label{subsec:DC=}

Let us consider $\mb{R}_i(t,s)$ defined in Eq.~(\ref{R=}); 
  or $\mb{R}_i(t) = \mb{r}_i(t) - \mb{r}_i(0)$, 
  which suffices if the system is statistically steady.
We denote the Cartesian components of $\mb{R}_i$, 
  in the 2D setup,
  with 
\begin{equation}
  \mb{R}_i
  = \TwoVect{{R}_{i1}}{{R}_{i2}}
  \label{R@}, 
\end{equation}
  omitting the time arguments when obvious.
The tensorial product of $\mb{R}_i$ and $\mb{R}_j$
  has four components, 
  which could be expressed in a matrix form, 
  as 
\begin{equation}
  \mb{R}_i\otimes\mb{R}_j 
  = 
  \begin{bmatrix}
    {{R}_{i1}}{{R}_{j1}}  &
    {{R}_{i1}}{{R}_{j2}}  \\
    {{R}_{i2}}{{R}_{j1}}  &
    {{R}_{i2}}{{R}_{j2}}  
  \end{bmatrix} 
  \label{RR@}. 
\end{equation}

Our target 
  is the \emph{displacement correlation tensor}, 
\begin{equation}
  \ChiR(\di,t) 
  = \Av{\mb{R}\otimes\mb{R}}_{\di}
  = 
  \begin{bmatrix}
    \Av{{R_1}{R_1}}_{\di} & \Av{{R_1}{R_2}}_{\di} \\
    \Av{{R_2}{R_1}}_{\di} & \Av{{R_2}{R_2}}_{\di}
  \end{bmatrix}
  \label{RR.2*2},
\end{equation}
  where $\Av{\quad}_{\di}$
  denotes conditional averaging 
  over the pairs of particles $(i,j)$
  whose relative position vector at the initial time
  equals $\di\,(\ne\mb{0})$.
If the initial time is taken at $t=0$, 
  the condition is expressed as follows:
  the initial value of the relative position vector, 
  $\rInit_{ij} = \mb{r}_j(0) - \mb{r}_i(0)$,  
  should be equal to $\di$.
The symbol $\ChiR(\di,t)$ with the arguments $(\di,t)$ 
  is used 
  in order to make it clear 
  that the displacement correlation tensor 
  in Eq.~(\ref{RR.2*2})
  is a function of the initial distance $\di$ 
  and the time interval $t$.  
The conditional averaging 
  can be re\-expressed as \cite{note.ChiR}
\begin{equation}
  \ChiR(\di,t) 
  =
  \frac{L^\nd}{N^2}
  \Av{%
  \sum_{i,j}
  \frac{%
  \delta^\nd( \rInit_{ij} - \di )
  }{g_2(\rInit_{ij})}
  \mb{R}_i\otimes\mb{R}_j
  }
  \label{RR.nD}, 
\end{equation}
  where 
\[
  g_2(\mb{r}) 
  = 
  \frac{L^\nd}{N^2}
  \Av{%
  \sum_{i',j'} 
  \delta^\nd(\mb{r}_{j'}-\mb{r}_{i'}-\mb{r})
  }\quad 
  (\mb{r}\ne\mb{0})\relax.
\]
Note that $\ChiR(\di,t)$ is undefined 
  for $\abs{\smash{\di}\mathstrut}<\sigma$
  because the denominator in Eq.~(\ref{RR.nD}) vanishes.
In the 1D setup,
  the formulation of the problem can be simpler
  \cite{Ooshida.PRE88,Ooshida.MPLB29}.
Since we can take it for granted 
  that the particles are numbered consecutively
  if $\Vmax \gg \kT$ \cite{note.number},
  it is meaningful 
  to calculate the displacement correlation
  in the form of $\Av{{R_i}{R_j}}$ 
  as a function of $t$ and $|j-i|$.
In higher dimensions,
  of course,
  the meaning of $\Av{{\mb{R}_i}\otimes{\mb{R}_j}}$ is not clear 
  unless the way of numbering is specified,
  and therefore $\ChiR(\di,t)$ must be used instead.

For the sake of simplicity,
  when the distinction in regard to the independent variable 
  is not important,
  we may denote both $\ChiR(\di,t)$ and $\Av{{R_i}{R_j}}$
  symbolically with $\Av{\mb{R}\otimes\mb{R}}$.

\subsection{Numerical demonstration of vortical motion}
\label{subsec:num}

\begin{figure}  
  \IncFig[clip,width=7.0cm]{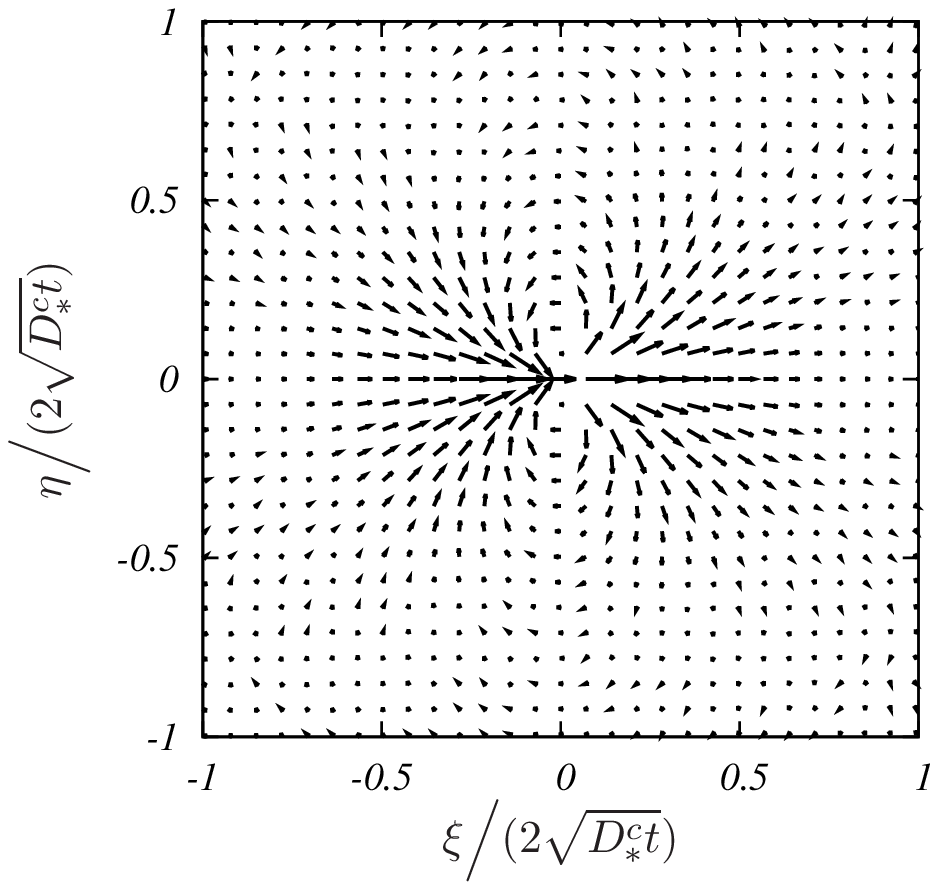}
  \caption{\label{Fig:RR.2D}%
    Displacement correlation in the 2D system ($\phi=0.5$). 
    The numerical value of $\ChiR(\di,t) \cdot \mb{e}_1$, 
    computed for $t = 0.8$, 
    is plotted as a vector field on the $\di$-plane.
  }
\end{figure}

As a pictorial representation 
  of the vortical cooperative motion,  
  we plot the numerical value of  
\[  
  \ChiR(\di,t) \cdot \mb{e}_1
  = \TwoVect{\Av{{R_1}{R_1}}_{\di}}{\Av{{R_2}{R_1}}_{\di}}
\]  
  as a vector field on the $\di$-plane, 
  where $\mb{e}_1$ is the $x$-directional unit vector.
The result 
  is shown in Fig.~\ref{Fig:RR.2D}.  
It demonstrates 
  basically the same flow structure 
  as in Fig.~8 of Doliwa and Heuer \cite{Doliwa.PRE61}, 
  with a pair of vortices.

The displacement correlation is computed 
  on the assumption 
  that the system is statistically steady, 
  homogeneous, and isotropic, 
  and also has the reflectional symmetry.
Due to these symmetries (as is 
  well-known in the theory of homogeneous turbulence
  \cite{Batchelor.Book1953,Hinze.Book1975,Landau.fluid}),
  the displacement correlation tensor must be expressible 
  in the form 
\begin{equation}
  \ChiR(\di,t)
  = \Xl (\diScalar/\ell_0,t) \frac{\di\otimes\di}{\di^2}
  + \Xtr(\diScalar/\ell_0,t)
  \left( \openone - \frac{\di\otimes\di}{\di^2} \right)
  \label{Xl+Xtr},
\end{equation}
  where $\diScalar = \abs{\smash{\di}}$.
The longitudinal and transverse correlations
  are denoted with $\Xl$ and $\Xtr$, respectively.
The positional arguments of these functions 
  are scaled with $\ell_0$ 
  for later convenience.
In the numerical calculation,  
  $\Xl$ and $\Xtr$ are evaluated
  by decomposing $\mb{R}_i$ and $\mb{R}_j$ 
  in Eq.~(\ref{RR.nD}) 
  into the components parallel and perpendicular to $\rInit_{ij}$
  and then averaging their products
  with a method similar to the one 
  explained in the Appendix A of Ref.~\cite{Ooshida.PRE88}.
The average was typically taken 
  over 9600 samples with independent initial conditions.

We emphasize 
  that the displacement correlation, 
  $\Av{\mb{R}\otimes\mb{R}}$,  
  is a two-time statistical quantity 
  involving four points in the space--time. 
One may question 
  whether such a complicated quantity is necessary, 
  because it seems possible to capture vortical motions 
  by means of one-time correlation of the velocity field, 
  as is the case in fluid or granular turbulence 
  \cite{van-Noije.PRE61,Frisch.Book1995}.
Unfortunately, 
  due to the Langevin noise  
  that breaks the momentum conservation, 
  the velocity field in the colloidal liquid is so noisy
  that its simple spatial correlation 
  does not provide useful information.
In other words, 
  the Langevin noise 
  disqualifies the momentum (and the velocity)
  for treatment as a slow variable; 
  this is why a two-time quantity $\mb{R}$ 
  must be used instead of the instantaneous velocity.

One may notice, by careful observation,  
  a subtle difference 
  between Fig.~\ref{Fig:RR.2D} in the present paper 
  and Fig.~8 of Doliwa and Heuer \cite{Doliwa.PRE61}: 
  the former is symmetric in regard to reflection 
  along the $x$-axis, 
  while the latter appears somewhat asymmetric.
This disagreement seems to have originated 
  from the different ways to incorporate the directionality
  of the ``central'' particle.
Doliwa and Heuer \cite{Doliwa.PRE61}
  produced their Fig.~8 
  by averaging the direction of $\mb{R}_j$
  for all the pairs $(i,j)$,  
  on the condition that the whole system is turned 
  so that $\mb{R}_i$ points 
  in the \emph{positive} direction of the $x$-axis.
On the other hand,  
  Eq.~(\ref{Xl+Xtr}) is invariant 
  under replacement of $\di$ with $-\di$,  
  which means that our definition incorporates 
  displacements 
  both \emph{parallel} and \emph{antiparallel} to $\di$
  with equal weightings. 
In this sense,  
  due to the present definition of $\Av{\mb{R}\otimes\mb{R}}$
  which is simpler 
  than the procedure in Ref.~\cite{Doliwa.PRE61},
  some information might be lost.
Nevertheless, 
  the present definition is advantageous 
  to analytical calculation, 
  as will be seen in the next section.
The asymmetry would be captured 
  by a higher-order correlation 
  such as $\Av{\abs{\mb{R}_i-\mb{R}_j}^2 (\mb{R}_i-\mb{R}_j)}$, 
  which we leave as a future work.

\section{Analytical calculation of displacement correlation tensor}
\label{sec:theory}

In what follows,  
  we outline the procedure 
  for analytical calculation of $\Av{\mb{R}\otimes\mb{R}}$,  
  developed by the group of the present authors 
  \cite{Ooshida.PRE88,Ooshida.MPLB29,Ooshida.BRL11} 
  with the label variable method 
  \cite{Ooshida.JPSJ80}.
The idea is clarified 
  in the 1D setup 
  and then applied to the 2D case.

\subsection{One-dimensional theory}

As was pointed out 
  by Doliwa and Heuer themselves \cite{Doliwa.PRE61},  
  the phenomenon of a growing dynamical length scale 
  is exhibited by simpler systems, 
  like a 1D chain of $N$ diffusive particles 
  connected by harmonic springs, 
  as described by the Langevin equation
\begin{equation}
  \mu\dot{X}_i 
  = \kappa\left( X_{i+1} - 2 X_i + X_{i-1} \right) 
  + \mu f_i(t) 
  \label{Rouse}
\end{equation}
  with $\kappa$ denoting the effective spring constant
  \cite{Lizana.PRE81,Centres.PRE81}.
This is also known as the Rouse model 
  in the context of polymer dynamics 
  \cite{Doi.Book1986}.

For the system described by Eq.~(\ref{Rouse}), 
  where $X_i$ denotes the position of the $i$-th particle
  numbered consecutively,  
  the displacement correlation is obtained analytically 
  \cite{Majumdar.PhysicaA177,Ooshida.PRE88}.
Taking the continuous limit 
  for simplicity,  
  we rewrite Eq.~(\ref{Rouse}) 
  in terms of $h = h(\xi,t) = x(\xi,t) - \ell_0\xi$, 
  with $\xi$ here denoting a continuum analogue of $i$,
  and with $\ell_0$ defined in Eq.~(\ref{len0}).
The equation for $h$ 
  reads
\begin{equation}
  \dt{h(\xi,t)} 
  = D' \dXi^2{h(\xi,t)} + f_h(\xi,t) 
  \label{EW}, 
\end{equation}
  where $D' = \kappa/\mu$, and 
  $f_h$ is a thermal noise such that
\[
  \Av{{f_h(\xi,t)}{f_h(\xi',t')}} 
  = 2D \delta(\xi-\xi') \delta(t-t')
  \relax.
\]
Since Eq.~(\ref{EW}) 
  is readily solved in the Fourier representation, 
  various statistical quantities can be calculated analytically.
In particular, 
  by noticing 
  that the displacement of the particle labeled with $\xi$ 
  is given by 
\begin{equation}
  R(\xi,t,s) 
  = x(\xi,t) - x(\xi,s) 
  = h(\xi,t) - h(\xi,s) 
  \label{R.h},
\end{equation} 
  the displacement correlation 
  is obtained in the form 
\begin{equation}
  \frac{\Av{ R(\xi,t,s) R(\xi',t,s) }}{\sqrt{D'(t-s)}}
  \propto \varphi(\theta)
  \label{sim.1D}
\end{equation}
  in terms of the similarity variable 
  $\theta = (\xi-\xi')/\ld(t-s)$, 
  where $\ld(t-s) \propto \sqrt{t-s}$ 
  is the dynamical correlation length 
  that diverges to infinity for $t-s \to +\infty$.

Recently, 
  the group of the present authors 
  succeeded in extending the above analysis 
  to colloidal systems 
  modeled by the Dean--Kawasaki equation (\ref{DK.Q})
  \cite{Ooshida.PRE88,Ooshida.MPLB29,Ooshida.BRL11}. 
One of the key ideas 
  is to introduce the variable $\xi$, 
  referred to as the \emph{label variable}, 
  not through consecutive numbering of the particles
  but through the equation 
\begin{equation}
  (\rho,Q) =
  \begin{vmatrix}
     \mb{e}_0^{} & {\partial_t}\xi \\
     \mb{e}_1^{} & {\partial_x}\xi 
  \end{vmatrix}
  = ({\dx\xi}, \;{-\dt\xi})
  \label{L.1D},  
\end{equation}
  where $\mb{e}_0$ and $\mb{e}_1$ 
  are the unit vectors 
  along the $t$-axis and the $x$-axis 
  of the $1+1$-dimensional space--time.
It is readily shown,
  by taking the inner product of $(\rho,Q)$ in Eq.~(\ref{L.1D})
  and $(\dt\xi, \dx\xi)$, 
  that $\xi = \xi(x,t)$ satisfies the convective equation, 
\[
  \rho ( \dt + u \dx ) \xi(x,t) = 0
  \relax, 
\]
  where $u$ is the velocity field such that $Q = \rho u$.
It implies 
  that specifying a constant value of $\xi(x,t)$ 
  works as a label for the world\-line of a particle, 
  hence the name \emph{label variable}.
The mapping 
  from $x$ to $\xi$ is then inverted, as 
\begin{equation}
  \xi \mapsto x = x(\xi,t)
  \label{map.xi-to-x}, 
\end{equation}  
  so that $\xi$ is hereafter treated 
  as an independent variable.
Note that this inversion would be highly singular 
  if the delta function in Eq.~(\ref{rho=}) 
  were not blunted; 
  in actuality, 
  this singularity is mitigated 
  as a consequence of the coarse-graining.
Another crucial idea 
  consists in adoption of the field variable $\psi$, 
  as a link 
  between the density field 
  and the displacement correlation. 
The variable $\psi$
  is defined by the relation 
\begin{equation} 
  \dXi{x(\xi,t)} = \ell_0 \left[ 1+\psi(\xi,t) \right]
  \label{dx/dXi}; 
\end{equation}
  as the expression on the left-hand side 
  equals $1/\rho$
  due to Eq.~(\ref{L.1D}), 
  $\psi$ can be interpreted 
  as elongation of the particle interval 
  \cite{Ooshida.JPSJ80,Spohn.JStatPhys154}.
The adoption of $\psi$ as the field variable 
  corresponds to the vacancy-based description 
  of 1D lattice gases \cite{van-Beijeren.PRB28,Bakhti.PRE89}.
In terms of the Fourier representation of $\psi$,  
  defined as 
\begin{equation}
  \check\psi(k,t) 
  = {\frac1N} \int e^{\II k\xi} \psi(\xi,t) \D\xi
  \quad
  \left(  \frac{k}{2\pi/N} \in \Z  \right)
  \label{m1}   
\end{equation}
  so that 
\begin{equation}
  \psi(\xi,t) = \sum_k \check\psi(k,t) e^{-\II k\xi}
  \label{m2}, 
\end{equation}
  we consider the two-time correlation  
\begin{equation}
  C(k,t,s) = \frac{N}{L^2} \Av{\check\psi(k,t)\check\psi(-k,s)}
  \label{Corr=}; 
\end{equation}
  if the system is statistically steady,
  then it suffices to consider $C(k,t') = C(k,s+t',s)$.
Using the Fourier representation in Eq.~(\ref{m2}) 
  and the relation 
\begin{equation}
  \dXi{R(\xi,t,s)} 
  = \ell_0 \left[ \psi(\xi,t) - \psi(\xi,s) \right]
  \label{dR/dXi}
\end{equation}
  derived from Eq.~(\ref{dx/dXi}),
  we can construct a formula 
  that gives the displacement correlation 
  in terms of $C$ in Eq.~(\ref{Corr=}).
In the case of statistically steady systems, 
  the formula reads 
\begin{multline}
  \Av{R(\xi,t)R(\xi',t)}   \\ {}
  = \frac{L^4}{\pi N^2}  \int_{-\infty}^{\infty} 
  e^{-{\II}k(\xi-\xi')}  \frac{C(k,0) - C(k,t)}{k^2}
  \D{k} 
  \label{AP+},
\end{multline}%
  where $R(\xi,t) = x(\xi,t) - x(\xi,0)$;
  Eq.~(\ref{AP+}), as well as its extensions,  
  is referred to as the Alexander--Pincus formula, 
  named after the authors of a celebrated work 
  on single-file diffusion 
  \cite{Alexander.PRB18}.

We note that Eq.~(\ref{AP+}) can be inverted 
  into the form 
  that gives $[C(k,0)-C(k,t)]/k^2$
  in terms of the displacement correlation. 
This form will be discussed later in Subsec.~\ref{subsec:invAP},
  together with its two-dimensional analogues, 
  as the ``inverse'' of our present strategy:
  we are targeting at $\Av{\mb{R}\otimes\mb{R}}$ 
  by means of the Alexander--Pincus formula, 
  into which we need $C$ as an input.

In order to calculate $C$, 
  we rewrite
  the Dean--Kawasaki equation 
  by changing the independent variables 
  from $(x,t)$ to $(\xi,t)$ 
  \cite{Ooshida.JPSJ80,Ooshida.PRE88,note.chain}.
The continuity equation (\ref{+cont})
  is replaced by 
\begin{equation}
  \ell_0 \dt\psi(\xi,t) = \dXi{u(\xi,{t})}
  \label{cont.psi}
\end{equation}
  in the 1D case, 
  and we substitute $u = Q/\rho$ into Eq.~(\ref{cont.psi})
  where $Q$ is given by the 1D version of Eq.~(\ref{DK.Q}).
Switching to the Fourier representation in terms of $\check\psi$,  
  we have
\begin{equation}
  \partial_t  \check\psi(k,t)  
  = 
  -\Dc_* k^2 \check\psi(k,t) 
  + \Order(\check\psi^2) 
  + \rho_0 \check{f}_{\text{L}}(k,t)
  \label{d2psi}
\end{equation}
  with the statistics of the random force term
  given as 
\begin{equation}
  \rho_0^2 \Av{\CheckFL(k,t) \CheckFL(-k',t')}
  = \frac{2 D_*}{N} k^2 \delta_{kk'} \delta(t-t')  
  \label{f4};  
\end{equation}
  here we have defined 
  $ D_* = D/{\ell_0^2}$ 
  and 
  $ \Dc_* = \Dc/{\ell_0^2}$, 
  with $\Dc = \Dc(k)$ denoting 
  the collective diffusion coefficient 
  in Eq.~(\ref{Dc})
  at the label-space wavenumber $k$ 
  (physical wavelength ${2\pi}\ell_0/k$).
Within the linear approximation,
  Eq.~(\ref{d2psi}) 
  readily yields  
\begin{equation}
  C(k,t)
  = 
  \dfrac{S(k)}{L^2} e^{-{\Dc_*}k^2t}
  \label{d4psi},
\end{equation}
  which is substituted 
  into the Alexander--Pincus formula (\ref{AP+}).
Since the contribution from the long-wave modes 
  is dominant, 
  the $k$-dependence of $S(k) \simeq S(0) + \Order(k^2)$ 
  can be ignored, 
  and $\Dc_*$ is also regarded as independent of $k$.
Then, upon evaluation of the integral,
  we obtain 
\begin{equation}
 \frac{\Av{ R(\xi,t) R(\xi',t) }}{\sigma \sqrt{{\Dc}t}} 
 = \frac{2S}{{\rho_0}\sigma}
 \left( 
   \frac{{~}e^{-\theta^2}\!}{\sqrt{\pi}} - |\theta| \erfc|\theta| 
 \right)
 = \varphi(\theta)
 \label{R1*R2.sim},
\end{equation}
  in terms of a similarity variable,
  $\theta = (\xi-\xi')/\ld(t)$, 
  with the dynamical correlation length 
  $\ld(t) = 2\sqrt{{\Dc}t}$.
This is evidently in the form of Eq.~(\ref{sim.1D}).

\begin{figure}  
  \IncFig[clip,width=7.0cm]{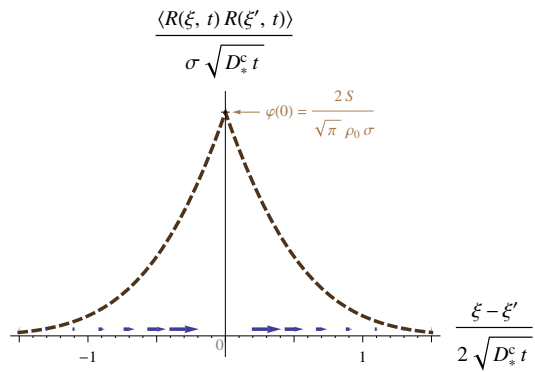}
  \caption{\label{Fig:RR.1D}%
    Displacement correlation in the 1D case,
    given in Eq.~(\ref{R1*R2.sim}).  
  }
\end{figure}

The function $\varphi$ in Eq.~(\ref{R1*R2.sim})
  is plotted in Fig.~\ref{Fig:RR.1D}.
To emphasize the vectorial character of the displacement, 
  the 1D vector field
\begin{equation}
  \frac{\Av{\mb{R}(\xi,t)\otimes\mb{R}(0,t)}}{\sigma\sqrt{{\Dc}t}}
  \cdot\mb{e}_1
  = \varphi\left(\frac{\xi}{\ld(t)}\right) \mb{e}_1
  \notag 
\end{equation}
  is shown together;
  this might be interpreted 
  as illustrating the average collective motion of the particles, 
  on the condition 
  that the central particle has moved rightward. 

A nonlinear theory can be developed 
  by taking into account the quadratic nonlinear terms
  that are ignored in Eq.~(\ref{d2psi}), 
  in the form of mode-coupling theory (MCT) for $C(k,t)$  
  \cite{Ooshida.PRE88}.
The notorious difficulty 
  concerning the violation of the fluctuation-dissipation theorem 
  is naturally resolved by the choice of the variable,
  namely $\psi(\xi,t)$ instead of $\rho(x,t)$.
Although there are 
  an infinite number of nonlinear terms 
  in the power series for the ideal solution entropy, 
  $\log(1+\psi) = \psi - \psi^2/2 + \cdots$, 
  we truncated the series focusing on the longtime behavior,
  with the exponential decay in Eq.~(\ref{d4psi}) in mind; 
  this treatment 
  is numerically justified 
  by asymptotic validity 
  of the linear result in Eq.~(\ref{R1*R2.sim}).
The memory integral in the MCT equation
  gives a correction to $C(k,t)$ in Eq.~(\ref{d4psi}),  
  resulting in modification of Eq.~(\ref{R1*R2.sim}) 
  as 
\begin{equation}
  \Av{R(\xi,t)R(\xi',t)}
  = 
  \sigma\sqrt{{\Dc}t}\,\varphi(\theta)
  - 
  \frac{\sqrt2}{3\pi}\ell_0^2
  \left( 1 - 2\theta^2 \right) e^{-\theta^2}
  \label{R1*R2.NL}.
\end{equation}  
The correction term is shown to have a perceptible effect 
  on the mean square displacement (MSD), i.e.\ 
  the short-length limit of the displacement correlation, 
  in regard to its finite-time behavior
  \cite{Ooshida.PRE88,Ooshida.BRL11,note.C0}.

The displacement correlation 
  can be regarded as generalization of MSD 
  to the two-particle statistics,
  and is related 
  to another form of two-particle two-time correlation, 
  $\Av{\left[{x(\xi,t) - x(\xi',0)}\right]^2}$ 
  in our notation.
Information of the inter-particle distance correlation 
  \cite{Lizana.PRE81}, 
  regard\-able as an analogue of bond breaking 
  \cite{Yamamoto.PRE58,Shiba.PRE86,Kawasaki.PRE87}, 
  is also contained in this form of correlation. 
Recently, 
  the probability distribution function 
  corresponding to this correlation 
  was shown to be calculable exactly 
  in the special case of the 1D system of point particles
  \cite{Sabhapandit.JStat2015}.
Instead of going into details of 1D systems, 
  however, 
  we will now focus on the problem 
  of how to proceed to the 2D case.

\subsection{Relation between the initial position and the label variable}
\label{subsec:di}

As was remarked  
  at the end of Subsec.~\ref{subsec:DC=},
  we have sometimes denoted the displacement correlation 
  symbolically as $\Av{\mb{R}\otimes\mb{R}}$, 
  without specifying the independent variable.
This is not a serious problem 
  when the consecutive numbering or its continuum limit 
  provides a natural choice of the independent variable, 
  as is the case in the 1D system.

Before proceeding to the 2D theory, 
  however,
  the relation between the label variable, $\xi$, 
  and the initial position of the particle, $\xInit = x(\xi,0)$, 
  should be briefly discussed.
In general, 
  there is a certain degree of freedom 
  in definition of the label variable.
At the hydrodynamic scale 
  where continuum description is valid in the usual sense, 
  the label variable should be a smooth one-to-one function 
  of the initial position.
A popular choice 
  is to select the identity function, so that $\xi = \xInit$, 
  or to define $\xi = \xInit/\ell_0$ 
  if $\xi$ is desired to be nondimensional.
However, we have not made  
  this naive choice.

Our definition of the label variable in Eq.~(\ref{L.1D})  
  is devised so as to reproduce the particle numbering 
  in the 1D setup.
It allows for the density fluctuation at $t=0$,
  as is seen from Eq.~(\ref{dx/dXi}) 
  which implies
\begin{equation}
  \xInit_{\mathrm{B}} - \xInit_{\mathrm{A}} 
  = \ell_0 \int_{\mathrm{A}}^{\mathrm{B}} 
  \left[ 1 + \psi(\xi,0) \right]\D\xi
  \label{di=int}
\end{equation}
  for the initial distance 
  between the particles $\mathrm{A}$ and $\mathrm{B}$.
Due to the presence of $\psi$
  on the right-hand side of Eq.~(\ref{di=int}), 
  $\xi_{\mathrm{B}} - \xi_{\mathrm{A}}$ 
  differs from 
  $(\xInit_{\mathrm{B}} - \xInit_{\mathrm{A}})/\ell_0$
  in general. 
However, 
  since the integral of $\psi$ tends to average out to zero 
  for large distance, 
  we may expect 
\begin{equation}
  \frac{\xInit_{\mathrm{B}} - \xInit_{\mathrm{A}}}{\ell_0} 
  \simeq \xi_{\mathrm{B}} - \xi_{\mathrm{A}} 
  \label{di=xi.1D}
\end{equation}   
  as an \emph{approximate} relation. 
This approximation
  also justifies assuming $C(k,0) \simeq S(0)/L^2$ for small $k$, 
  even if nonlinearity is taken into account.

If we accept Eq.~(\ref{di=xi.1D}),  
  the displacement correlation 
  as a function of $\diScalar$ and $t$, 
  $\ChiR(\diScalar,t) 
  = X_{\mathrm{1D}}(\diScalar,t)\mb{e}_1\otimes \mb{e}_1$, 
  is approximated as 
\begin{equation}
  X_{\mathrm{1D}}(\diScalar,t)
  \simeq 
  \Ev{\Av{{R(\xi,t)}{R(\xi',t)}}}{\xi=\xi'+\diScalar/\ell_0} 
  \label{Chi@di=RR@xi.1D}.
\end{equation} 
The validity of this approximation
  can be established
  by rewriting the 1D version of Eq.~(\ref{RR.nD})
  in the form of an integral over the initial position
  (see Subsec.~V-A in Ref.~\cite{Ooshida.PRE88}).
After some calculation, 
  $X_{\mathrm{1D}}(\tilde{d},t)$
   is found to be expressible 
  as a sum of two parts:
  the first part simply gives Eq.~(\ref{Chi@di=RR@xi.1D}),
  while the second part involves 
  triple correlations.
The contribution from these triple correlation terms 
  is shown to be asymptotically negligible for large $t$.
In this sense, 
  it is acceptable
  to identify the label distance $\xi-\xi'$ 
  with the initial distance $\diScalar$ divided by $\ell_0$, 
  at least as a crude approximation. 
The approximation should be better 
  for larger distance in space or time.

\subsection{Two-dimensional theory}

The procedure for calculation of $\Av{\mb{R}\otimes\mb{R}}$
  for 2D colloidal liquids \cite{Ooshida.BRL11}
  is essentially parallel to its 1D prototype. 
The starting point is the Dean--Kawasaki equation,  
  written in the form of Eq.~(\ref{DK.Q})
  combined with the continuity equation (\ref{+cont}).
Subsequently, 
  we introduce the label variable $\BoldXi = (\xi,\eta)$
  through the 2D analogue of Eq.~(\ref{L.1D}); 
  we define also the tensorial field variable $\Psi$, 
  which is essentially the deformation gradient tensor 
  \cite{Marsden.Book1994}, 
  by generalizing Eq.~(\ref{dx/dXi})
  to the 2D case. 
By calculating the two-time correlations 
  of the components of $\Psi$,
  in the Fourier representation, 
  and substituting them 
  into the 2D version of the Alexdander--Pincus formula, 
  we obtain $\ChiR(\di,t)$.

In comparison to the 1D case, 
  there are two main differences:
  firstly, various quantities appearing in the theory 
  are now tensorial, 
  and secondly, 
  we cannot depend on the consecutive numbering of the particles. 
For a pair of particles labelled with $\BoldXi$ and $\BoldXi'$,
  we expect
  ${\di}/{\ell_0} \simeq \BoldXi-\BoldXi'$ 
  for the initial relative position vector $\di$, 
  and therefore  
\begin{equation}
  \ChiR(\di,t)
  \simeq 
  \begin{bmatrix}
    \Av{\smash{{R_1(\di/\ell_0,t)}{R_1(\mb{0},t)}}} \!& 
    \Av{\smash{{R_1(\di/\ell_0,t)}{R_2(\mb{0},t)}}} \\[0.25ex]
    \Av{\smash{{R_2(\di/\ell_0,t)}{R_1(\mb{0},t)}}} \!& 
    \Av{\smash{{R_2(\di/\ell_0,t)}{R_2(\mb{0},t)}}} 
  \end{bmatrix}
  \label{Chi@di=RR@xi.2D}
\end{equation} 
  where we have chosen $\BoldXi' = \mb{0}$ 
  without loss of generality, 
  assuming translational invariance.
As we further assume 
  rotational and reflectional invariance of the system, 
  $\ChiR(\di,t)$ must be expressed 
  in the form of Eq.~(\ref{Xl+Xtr}) 
  as a sum of the longitudinal and the transverse components, 
  represented by two functions, 
  namely $\Xl(\abs{\BoldXi},t)$ and $\Xtr(\abs{\BoldXi},t)$.
The goal of the calculation 
  is to obtain analytical expressions for these two functions.


The procedure of the calculation 
  starts with introducing the label variable $\BoldXi = (\xi,\eta)$
  and thereby rewriting 
  the Dean--Kawasaki equation (\ref{DK.Q}).
Recalling that Eq.~(\ref{L.1D})
  relates $\xi$ to $(\rho,Q)$, 
  we introduce $\BoldXi = (\xi,\eta)$
  as a solution to the equation
\begin{equation}
  (\rho,\mb{Q}) =
  \begin{vmatrix}
     \mb{e}_0^{} & \dt\xi& \dt\eta \\
     \mb{e}_1^{} & \dx\xi& \dx\eta \\
     \mb{e}_2^{} & \dy\xi& \dy\eta 
  \end{vmatrix}
  \label{*L.2D}.
\end{equation}
Then $\BoldXi$
  is demonstrated to satisfy the convective equation 
  and thus qualified as the label variable.
Note that an extension 
  to the three-dimensional case 
  is straightforward
  and the cofactor expansion reproduces the equations
  suggested in Ref.~\cite{Ooshida.JPSJ80}.

The mapping from the label variable 
  to the position,
\begin{equation}
  \BoldXi = (\xi,\eta) \mapsto 
  \mb{r}(\BoldXi,t) = \TwoVect{x(\xi,\eta,t)}{y(\xi,\eta,t)}
  \label{map.xi-to-r}, 
\end{equation}
  specifies a time-dependent curvilinear coordinate system
  which is sometimes referred to 
  as the \emph{convected coordinate system}
  \cite{Bird.Book1987,Ooshida.PRE77}.
The time-derivative of $\mb{r}(\BoldXi,t)$
  gives the velocity, 
  $\mb{u} = \dt\mb{r}(\BoldXi,t)$. 
The ``spatial'' derivative 
  gives what is called 
  the deformation gradient tensor \cite{Marsden.Book1994}
  or the displacement gradient tensor \cite{Bird.Book1987}.
Thereby we define $\Psi$, 
  as 
\begin{equation}
   \begin{bmatrix}
     \dXi{x} & \dHa{x}\\
     \dXi{y} & \dHa{y} 
   \end{bmatrix}
   = 
   \ell_0 \left( \openone + \Psi \right)
   \label{dr/dXi},
\end{equation}
  which is intended 
  as the 2D generalization of Eq.~(\ref{dx/dXi}).
Out of the four components of $\Psi$,
  only two are independent. 
If we choose the diagonal components of $\Psi$
  to represent them, as
\begin{equation}
  \Psi 
  = 
  \begin{bmatrix}
      \Psi_1 & \dHa\dXi^{-1}\Psi_1  \\
      \dXi\dHa^{-1}\Psi_2 & \Psi_2 
  \end{bmatrix}
  \label{Psi.mx2}, 
\end{equation}
  and introduce their Fourier modes by 
\begin{equation}
  \TwoVect{\Psi_1(\BoldXi,t)}{\Psi_2(\BoldXi,t)} 
  = \sum_{\mb{k}} 
  \TwoVect{\check\Psi_1(\mb{k},t)}{\check\Psi_2(\mb{k},t)} 
  e^{-\II\kXi}
  \label{2D.m2}, 
\end{equation}
  the Dean--Kawasaki equation (\ref{DK.Q})
  is rewritten as 
\begin{multline}
  \dt
  \TwoVect{\check\Psi_1(\mb{k},t)}{\check\Psi_2(\mb{k},t)}
  = 
  -\Dc_*
  \begin{bmatrix}
    k_1^2 & k_1^2  \\[0.5ex]%
    k_2^2 & k_2^2  
  \end{bmatrix}%
  \TwoVect{\check\Psi_1(\mb{k},t)}{\check\Psi_2(\mb{k},t)} 
  + \Order(\Psi^2) \\ {}
  + 
  \ell_0^{-1} 
  \TwoVect{\check{f}_1(\mb{k},t)}{\check{f}_2(\mb{k},t)}
  \label{2D-lin.Psi.eqn}; 
\end{multline}
  the forcing statistics are given by
\begin{multline}
  \ell_0^{-2}
  \begin{bmatrix}
    \Av{{\check{f}_1(\mb{k},t)}{\check{f}_1(-\mb{k}',t')}}&
    \Av{{\check{f}_1(\mb{k},t)}{\check{f}_2(-\mb{k}',t')}}\\
    \Av{{\check{f}_2(\mb{k},t)}{\check{f}_1(-\mb{k}',t')}}&
    \Av{{\check{f}_2(\mb{k},t)}{\check{f}_2(-\mb{k}',t')}}    
  \end{bmatrix}
  \\ {}
  = 
  \frac{2{D_*}}{N}
  \begin{bmatrix}
    k_1^2 & 0     \\ 
    0     & k_2^2 
  \end{bmatrix}
  \delta_{\mb{k},\mb{k}'} \delta(t-t')
  \label{f.2D}, 
\end{multline}
  with $D_* = D/\ell_0^2 = \rho_0 D$.  
The factor $\delta_{\mb{k},\mb{k}'}$ 
  originates from the random distribution of the particles, 
  which should be uniform for small values of $\mb{k}$.

The linear homogenous part of Eq.~(\ref{2D-lin.Psi.eqn})
  has two eigenvalues: $-\Dc_*\mb{k}^2$ and $0$.
Accordingly, we diagonalize Eq.~(\ref{2D-lin.Psi.eqn})
  by re-expressing $({\check\Psi_1},\; {\check\Psi_2})$
  as a linear combination of the eigenvectors.
This diagonalization 
  corresponds to switching 
  to the dilatational and rotational modes,
  denoted with $\psiD$ and $\psiR$, respectively,
  and defined as
\begin{align}
  & \dXi{x} + \dHa{y} 
  = \ell_0
  \left( 2 + \sum_{\mb{k}} \psiD(\mb{k},t) e^{-\II\kXi} \right)
  \relax, \label{div.X}
  \\
  & \dXi{y} - \dHa{x}
  = \ell_0 \sum_{\mb{k}} \psiR(\mb{k},t) e^{-\II\kXi} 
  \label{rot.X}.
\end{align}
The correlations of these modes  
  are denoted by 
\begin{equation}
  C_{ab}(\mb{k},t,s)
  = 
  \frac{N}{L^4} 
  \Av{{\psi_a(\mb{k},t)}{\psi_b(-\mb{k},s)}}
  \label{CorrH=}
\end{equation}
  with $a, b \in \{\mathrm{d},\mathrm{r}\}$ and $s<t$. 
In the present case, however, 
  $\Crd$ is found to vanish, 
  and for the sake of brevity,
  we write $\Cd$ for $\Cdd$ and $\Cr$ for $\Crr$, 
  respectively.
The linearized equation 
  yields
\begin{subequations}%
\begin{align}
  \Cd(\mb{k},t,s)
  &= 
  \frac{S}{L^4} e^{-\Dc_*\mb{k}^2(t-s)}  
  \label{Lin2.Cd},  \qquad
  \\
  \Cr(\mb{k},t,s)
  &= 
  \frac{2D_*\mb{k}^2}{L^4} (s - o)
  \label{Lin2.Cr},
\end{align}%
\label{eqs:Lin2}%
\end{subequations}%
  where $o$ is a constant of integration, 
  interpretable as the time at which $\Cr$ is reset. 
Note that both $\Cd$ and $\Cr$ are independent 
  of the direction of $\mb{k}$.


From $\Cd$ and $\Cr$ in Eqs.~(\ref{eqs:Lin2}),  
  the displacement correlation is obtained 
  via the Alexander--Pincus formula \cite{Ooshida.BRL11}, 
  which now reads
\begin{widetext}%
\begin{align}
  &\relax
  \begin{bmatrix}
    \Av{{R_1(\BoldXi,t,s)}{R_1(\mb{0},t,s)}} &
    \Av{{R_1(\BoldXi,t,s)}{R_2(\mb{0},t,s)}} \\
    \Av{{R_2(\BoldXi,t,s)}{R_1(\mb{0},t,s)}} &
    \Av{{R_2(\BoldXi,t,s)}{R_2(\mb{0},t,s)}}   
  \end{bmatrix}
  \notag\\&{}
  = \frac{L^6}{{2\pi^2}N}
  \iint
  \left[ 
  \frac{\Cd(\mb{k},s,s) + \Cd(\mb{k},t,t)}{2} - \Cd(\mb{k},t,s)
  \right]
  \begin{bmatrix}
    k_1^2      & {k_1}{k_2} \\
    {k_2}{k_1} & k_2^2
  \end{bmatrix}
  \frac{e^{-\II\kXi}}{\mb{k}^4}  \D{k_1}\D{k_2}
  \notag\\&{}
  +  \frac{L^6}{{2\pi^2}N}
  \iint
  \left[ 
  \frac{\Cr(\mb{k},s,s) + \Cr(\mb{k},t,t)}{2} - \Cr(\mb{k},t,s)
  \right]
  \begin{bmatrix}
    k_2^2       & \!{-{k_1}{k_2}} \\
    -{k_2}{k_1} &      k_1^2
  \end{bmatrix}
  \frac{e^{-\II\kXi}}{\mb{k}^4}  \D{k_1}\D{k_2}
  \label{AP2.H}.
\end{align}%
\end{widetext}%
Equation (\ref{Lin2.Cd}) 
  gives 
\begin{multline}
  \frac{\Cd(\mb{k},s,s) + \Cd(\mb{k},t,t)}{2} - \Cd(\mb{k},t,s)
  \\ {} = 
  \frac{S}{L^4} \left[ 1 - e^{-\Dc_*\mb{k}^2(t-s)} \right]
  \label{Cd.AP}
\end{multline}
  for the integrand in the first term 
  on the right-hand side of Eq.~(\ref{AP2.H}), 
  while the corresponding expression in the second term 
  is calculated as  
\begin{equation}
  \frac{\Cr(\mb{k},s,s) + \Cr(\mb{k},t,t)}{2} - \Cr(\mb{k},t,s)
  = \frac{D_*\mb{k}^2}{L^4}(t-s)
  \label{Cr.AP}
\end{equation}
  from Eq.~(\ref{Lin2.Cr}).
The integrals are then evaluated 
  and the result is equated 
  to $\ChiR(\di,t)$ in Eq.~(\ref{Xl+Xtr}), 
  which yields
\begin{subequations}
\begin{align}
  \Xl(\abs\BoldXi,t) 
  &= 
  \frac{S}{{4\pi}\rho_0}
  \left[
  E_1({\theta^2}) + \frac{e^{-\theta^2}}{\theta^2}
  \right]
  \label{Xl}, 
  \\
  \Xtr(\abs\BoldXi,t) 
  &= 
  \frac{S}{{4\pi}\rho_0}
  \left[
  E_1({\theta^2}) - \frac{e^{-\theta^2}}{\theta^2}
  \right]
  \label{Xtr}
  \relax,
\end{align}%
\label{eqs:ChiR.L}%
\end{subequations}
  where $\BoldXi = (\xi,\eta) \simeq \di/\ell_0$ and
\begin{equation}
  \theta^2 = \frac{\xi^2 + \eta^2}{4{\Dc_*}t} 
  \simeq \frac{\di^2}{[\ld(t)]^2}
  \label{2D.th=}
\end{equation}
  with $\ld(t) = 2\sqrt{{\Dc}t}$,
  and 
  $E_1(\;\cdot\;)$ denotes the exponential integral
  \cite{Arfken.Book2013},
\begin{equation}
  E_1(z) 
  = \int_z^\infty \frac{e^{-\zeta}}{\zeta}\D{\zeta} 
  \label{expInt}.
\end{equation}
Using the asymptotic expansion of $E_1(z)$ for large $z$, 
  from Eqs.~(\ref{eqs:ChiR.L})
  we find
\begin{subequations}
\begin{align}
  \Xl(\abs\BoldXi,t) 
  &\simeq 
  \frac{S}{{4\pi}\rho_0} e^{-\theta^2}
  \left( 2 \theta^{-2} - \theta^{-4} + \cdots \right)
  \label{Xl.far}
  \\
  \Xtr(\abs\BoldXi,t) 
  &\simeq 
  \frac{S}{{4\pi}\rho_0} e^{-\theta^2}
  \left( - \theta^{-4} + \cdots \right)
  \label{Xtr.far}
\end{align}%
\label{eqs:ChiR.far}%
\end{subequations}
  for large $\theta$.


\subsection{Inverse of Alexander--Pincus formulae}
\label{subsec:invAP}

Our analytical calculation,
  starting from the Dean--Kawasaki equation, 
  is routed through the Alexander--Pincus formula, 
  into which we input $C$ in Eq.~(\ref{Corr=}) 
  or its 2D analogues in Eq.~(\ref{CorrH=})  
  to obtain $\Av{\mb{R}\otimes\mb{R}}$ as the output.
It is, however, sometimes convenient 
  to reverse a part of this course.

In the 1D case,  
  the Fourier inversion 
  of the Alexander--Pincus formula (\ref{AP+}) gives 
\begin{align}
  \frac{C(k,0) - C(k,t)}{k^2}
  &= 
  \frac{\rho_0^2}{2L^2} 
  \int e^{{\II}k\xi} \Av{R(\xi,t)R(0,t)} \D{\xi}
  \notag\\
  &= 
  \frac{\rho_0^2}{2{L^2}N}
  \sum_{i,j} \Av{{R_i}{R_j}} e^{{\II}k(j-i)} 
  \label{inv.AP+}, 
\end{align}
  where we have used the relation 
  $\D\xi = \rho\,\D{x}$.
Here the data set of the displacement $R_i$ 
  is taken as the input.
This form may be convenient 
  when one wishes to analyze the data of displacements 
  from experimental observation of a 1D system of particles 
  or from direct numerical simulation of such a system.

The inversion of Eq.~(\ref{AP2.H})
  is slightly more complicated,  
  as we need to separate $\Cd$ and $\Cr$.
The inverse formulae 
  read 
\begin{widetext}%
\begin{align}
  \frac{1}{k^2} \left[  
    \frac{\Cd(\mb{k},s,s) + \Cd(\mb{k},t,t)}{2} - \Cd(\mb{k},t,t)
  \right]
  &= 
  \frac{\rho_0}{2L^4} \iint 
  e^{\II\kXi}
  \Av{{R^{(\para\mb{k})}(\BoldXi,t,s)}{R^{(\para\mb{k})}(\mb{0},t,s)}}
  \D\xi\D\eta
  \label{inv.AP2.Cd},
  \\
  \frac{1}{k^2} \left[  
    \frac{\Cr(\mb{k},s,s) + \Cr(\mb{k},t,t)}{2} - \Cr(\mb{k},t,t)
  \right]
  &= 
  \frac{\rho_0}{2L^4} \iint 
  e^{\II\kXi}
  \Av{{R^{(\perp\mb{k})}(\BoldXi,t,s)}{R^{(\perp\mb{k})}(\mb{0},t,s)}}
  \D\xi\D\eta
  \label{inv.AP2.Cr}, 
\end{align}
\end{widetext}%
  where we have defined
\[
  R^{(\para\mb{k})} = \ek \cdot \mb{R}
  \relax, \quad 
  R^{(\perp\mb{k})} = \det\left( \ek, \mb{R} \right)
\]
  in terms of $\ek = \mb{k}/k$.

The expression 
  on the right-hand side of Eq.~(\ref{inv.AP2.Cr}) 
  is essentially identical to the quantity 
  studied by Flenner and Szamel \cite{Flenner.PRL114}.
In our notation, 
  this quantity is defined as 
\begin{equation}
  S_4^\perp(\mb{q},t) 
  = {\frac1N} \Av{\sum_{i,j} 
  {R^{(\perp\mb{q})}_i}(t) {R^{(\perp\mb{q})}_j}(t) 
  e^{{\II}\mb{q}\cdot\rInit_{ij}}}
  \label{S4.Tr},
\end{equation}
  where $\mb{q}$ 
  is Fourier-conjugate to $\rInit$.
For the length scales much greater than $\ell_0$,  
  the approximation
\[
  \mb{k}\cdot \BoldXi \simeq   \mb{q}\cdot\rInit_{ij} 
  \relax, \quad 
  \mb{k} = \ell_0\mb{q}  
\]
  is valid, 
  which makes the integral 
  on the right-hand side of Eq.~(\ref{inv.AP2.Cr}) 
  equal to $S_4^\perp(\mb{q},t)$.
Similarly, 
  one may define 
\begin{equation}
  S_4^\para(\mb{q},t) 
  = {\frac1N} \Av{\sum_{i,j} 
  {R^{(\para\mb{q})}_i}(t) {R^{(\para\mb{q})}_j}(t) 
  e^{{\II}\mb{q}\cdot\rInit_{ij}}}
  \label{S4.L}
\end{equation}
  and find it equal to the integral 
  on the right-hand side of Eq.~(\ref{inv.AP2.Cd}).

By comparing Eqs.~(\ref{inv.AP2.Cd})--(\ref{S4.L}),  
  we notice that $\Cr$ and $S_4^\perp$ 
  contain essentially the same information, 
  and also that $\Cd$ and $S_4^\para$ are equivalent.
The displacement correlation $\ChiR$ 
  contains information of both $\Cr$ and $\Cd$.
The relation of these correlations  
  to the two-time correlation of density, 
  given by $F(k,t)$ in Eq.~(\ref{F=}),    
  deserves a comment:
  $F$ represents the same dilatational motion as $\Cd$ in effect,
  but not the rotational motion.
This implies that, 
  while the first integral in Eq.~(\ref{AP2.H}) 
  might be expressible in terms of $F$ 
  within some approximation, 
  the second integral requires information of $\Cr$
  which is not reducible to $F$.

In regard to $\Xl$ and $\Xtr$ in Eq.~(\ref{Xl+Xtr}),  
  it should be warned
  that Fourier transforms of $\Xl$ and $\Xtr$ 
  differ from $S_4^\para$ and $S_4^\perp$.
This is naturally understood 
  once the difference between the two types of decomposition 
  are recognized:
  Eq.~(\ref{Xl+Xtr}) 
  is based on the orthogonal projection onto $\di$,  
  while the quantities in Eqs.~(\ref{inv.AP2.Cd})--(\ref{S4.L}) 
  are defined in terms of projection onto the wavenumber vector.

\section{Comparison of numerical and analytical results}
\label{sec:compare}

\begin{figure}
  \centering
  \IncFig{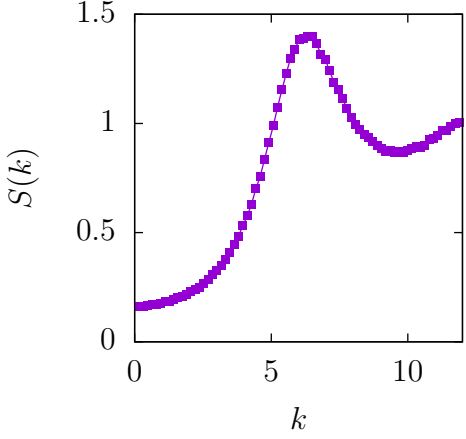}
  \caption{\label{Fig:Sk}%
  The static structure factor $S(k)$, 
  computed for $\phi=0.5$
  and giving $S = S(k\to+0) = 0.16$
  in this case.}
\end{figure}
\begin{figure}
  \centering
  \IncFig{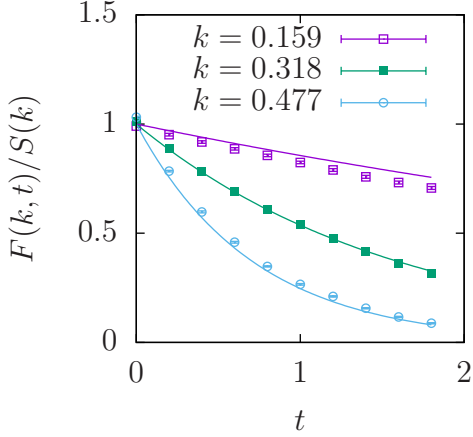}
  \caption{\label{Fig:Fkt}%
  Temporal decay of $F(k,t)$ for three values of $k$.
  Numerical results are shown with points 
  and compared with Eq.~(\ref{Dc}) plotted with the solid lines.}
\end{figure}

With the numerical data of 2D displacement correlation 
  illustrated in Fig.~\ref{Fig:RR.2D}
  and the analytical expressions 
  of  $\Xl (\abs\BoldXi,t)$ 
  and $\Xtr(\abs\BoldXi,t)$ 
  in Sec.~\ref{sec:theory},
  now let us discuss the validity 
  of the analytical results in Eqs.~(\ref{eqs:ChiR.L})
  by comparing them with the numerically computed values.
As was stated in Sec.~\ref{sec:setup}
  in regard to the area fraction $\phi$, 
  we focus on the case with $\phi = 0.5$;
  later, in Sec.~\ref{sec:discuss}, 
  the cases of several other values of $\phi$ 
  will be discussed briefly.
For the sake of simplicity,  
  hereafter we write $\xi$ instead of $\abs\BoldXi$, 
  when it is not confusing.

As the evaluation of the analytical expressions 
  in Eqs.~(\ref{eqs:ChiR.L}) 
  requires knowledge of the values 
  of $S = S(k\to+0)$ and $\Dc = \Dc(k\to+0)$, 
  we begin by computing these values.
The static structure factor $S(k)$ for $\phi = 0.5$,  
  computed according to the procedure in Ref.~\cite{Ikeda.PRE92}, 
  is plotted in Fig.~\ref{Fig:Sk}.
From this data, 
  we obtain 
  $S = S(k \to +0) = 0.16$
  by extrapolation,
  and incidentally, 
  we also find the peak wave\-number 
  $k_{\text{peak}} = 6.3\,\sigma^{-1}$.
Numerical values of $F(k,t)$, 
  defined by Eq.~(\ref{F=}), 
  are computed analogously 
  and plotted in Fig.~\ref{Fig:Fkt}, 
  together with the values from Eq.~(\ref{Dc})
  represented by the solid lines.  
As these values in Fig.~\ref{Fig:Fkt}  
  are seen to be consistent, 
  $\Dc$ is confirmed to be related to $S$ by Eq.~(\ref{Dc}).

\begin{figure}
  \centering
  \parbox{0.90\linewidth}{\SubFig{a}
  \IncFig{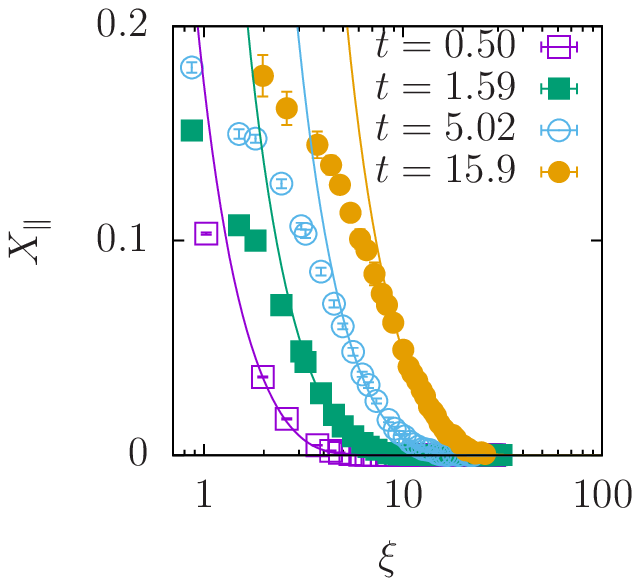}}\\[0.5\baselineskip]%
  \parbox{0.90\linewidth}{\SubFig{b}
  \IncFig{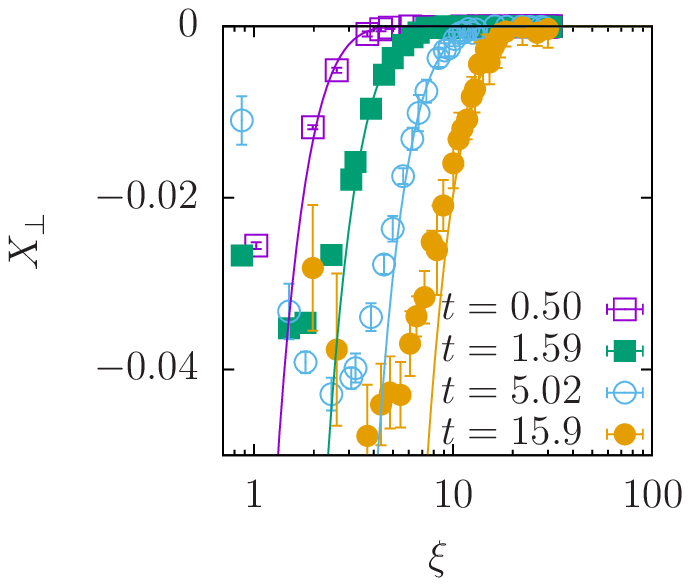}}
  \caption{\label{Fig:Chi}%
  Numerical values 
  of  the longitudinal correlation $\Xl (\xi,t)$ 
  and the transverse   correlation $\Xtr(\xi,t)$, 
  computed for four different time intervals 
  and plotted 
  against $\xi = \abs\BoldXi = \diScalar/\ell_0$.
  The predictions of Eqs.~(\ref{eqs:ChiR.L}) 
  are also shown with thin solid lines.}
\end{figure}
\begin{figure}
  \centering
  \parbox{0.90\linewidth}{\SubFig{a}
  \IncFig{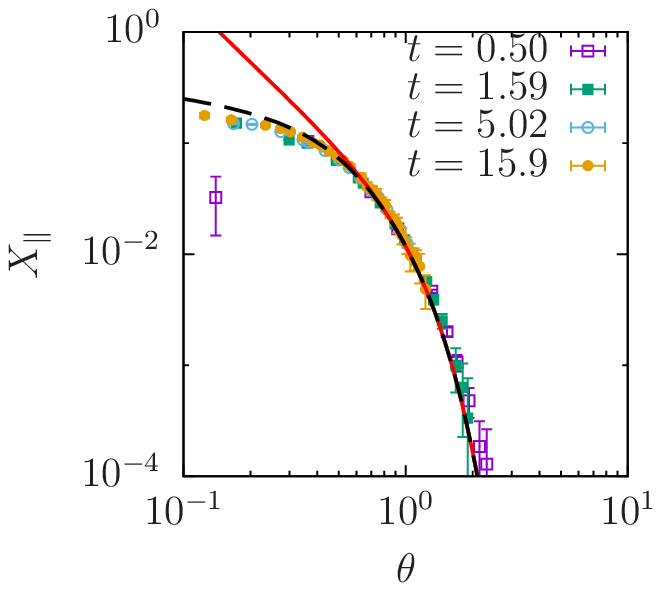}}\\[0.5\baselineskip]%
  \parbox{0.90\linewidth}{\SubFig{b}
  \IncFig{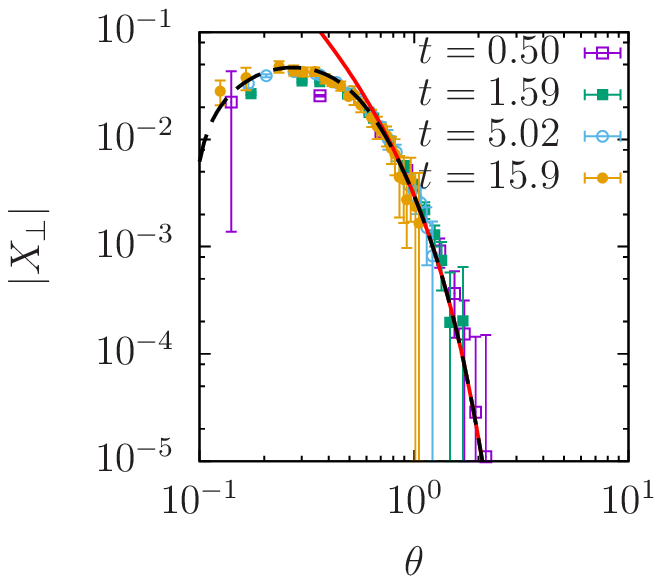}}
  \caption{\label{Fig:scale}%
  Verification of Eqs.~(\ref{eqs:ChiR.L})
  by plotting the longitudinal correlation $\Xl$ 
  and the transverse correlation $\Xtr$
  against the similarity variable  
  $\theta = \ell_0\abs\BoldXi/\ld(t) 
  = \abs\BoldXi/(2\sqrt{{\Dc_*}t}\,)$.
  The solid (red) lines 
  show theoretical predictions in Eqs.~(\ref{eqs:ChiR.L}),  
  while fitting with Eq.~(\ref{eqs:ChiR.mu})
  is delineated with broken (black) lines. }
\end{figure}

With the values of $S$ and $\Dc$ 
  thus obtained,
  we can calculate $\Xl$ and $\Xtr$ 
  according to the theoretical predictions in Eqs.~(\ref{eqs:ChiR.L})
  and compare them 
  with the corresponding results
  of direct numerical simulations.
Such comparison 
  is made in Fig.~\ref{Fig:Chi}, 
  where the functions $\Xl(\xi,t)$ and $\Xtr(\xi,t)$
  are plotted 
  against $\xi = \abs\BoldXi = \diScalar/\ell_0$.
Each panel contains 
  four sets of numerical data,
  corresponding to four different values 
  of the time difference $t$
  nondimensionalized with $\sigma^2/D$.
The thin solid lines, 
  representing the predictions of Eqs.~(\ref{eqs:ChiR.L}), 
  seem to agree with the computed results  
  for larger values of $\xi$, 
  i.e.\ 
  at large distances.
As $\xi$ increases,
  both $\Xl(\xi,t)$ and $\Xtr(\xi,t)$ tend to zero, 
  and the numerical data suggests 
  that $\abs{\Xtr}$ is always smaller than $\Xl$
  and seems to vanish faster than $\Xl$
  (notice the difference in the scale of the vertical axes);
  this is consistent 
  with Eqs.~(\ref{eqs:ChiR.L}), 
  implying $\abs{\Xtr} < \Xl$ for all finite values of $\xi$, 
  and also with the asymptotic behavior 
  in Eqs.~(\ref{eqs:ChiR.far}),
  which can be read as 
\begin{equation}
  \abs{\frac{\Xtr}{\Xl}} 
  \simeq {\frac12}\theta^{-2} = \frac{[\ld(t)]^2}{2\xi^2} \ll 1 
\end{equation}
  for large $\xi$.

A striking prediction of Eqs.~(\ref{eqs:ChiR.L})
  that $\Xl(\xi,t)$ and $\Xtr(\xi,t)$ 
  will collapse onto master curves,  
  if they are plotted against $\theta = \diScalar/\ld(t)$,
  is verified in Fig.~\ref{Fig:scale}.
From the plot of $\Xl$ 
  in Fig.~\ref{Fig:scale}(a), 
  it is seen 
  that all the four sets of numerical data 
  corresponding to the separate curves in Fig.~\ref{Fig:Chi}(a) 
  collapse onto a single master curve, 
  and Eq.~(\ref{Xl}), 
  plotted with a solid (red) line, 
  gives this master curve for $\theta\gtrsim1$.
Besides, 
  the master curve continues to the range of small $\theta$
  where Eq.~(\ref{Xl}) is not valid.
Analogously, 
  the four data sets in Fig.~\ref{Fig:Chi}(b) 
  collapse onto a master curve in Fig.~\ref{Fig:scale}(b)
  where $\abs{\Xtr}$ is plotted logarithmically 
  against $\theta$,
  and the master curve 
  agrees with Eq.~(\ref{Xtr}) for large $\theta$
  but deviates from it for small $\theta$.

Thus Eqs.~(\ref{eqs:ChiR.L}) is validated
  for moderate and large values of $\theta$. 
Besides, 
  at least in the case of $\phi = 0.5$, 
  $\Xl$ and $\Xtr$ are found to be  
  functions of $\theta$ alone in the entire range, 
  even if $\theta$ is so small that Eqs.~(\ref{eqs:ChiR.L}) fail.


\section{Discussion}
\label{sec:discuss}
\marginpar{}

\begin{figure}
  \centering
  \parbox{0.85\linewidth}{\SubFig{a}
  \IncFig[clip,width=6.5cm]{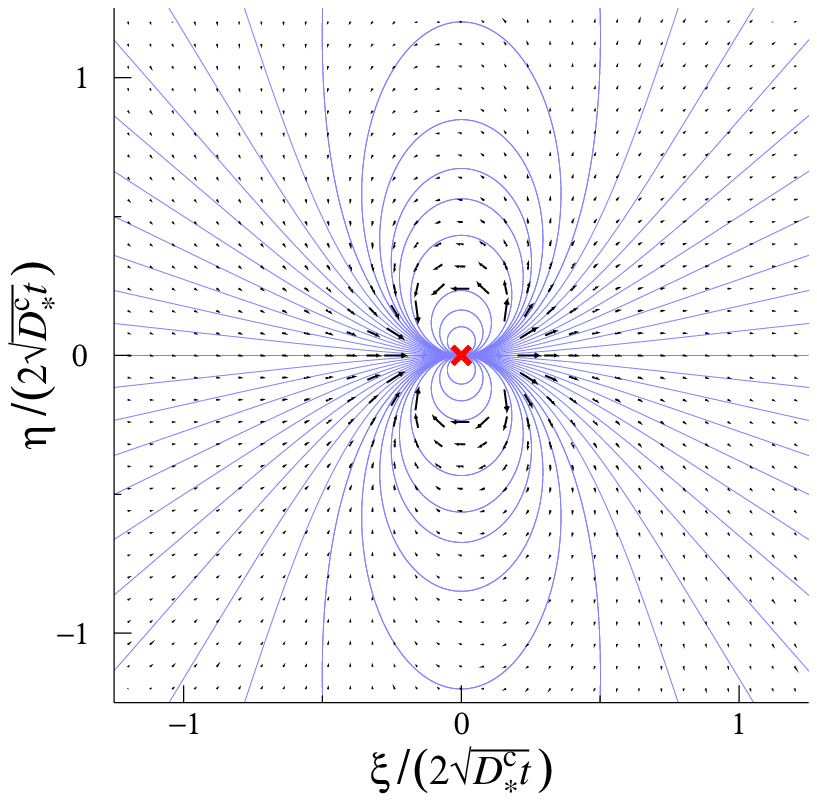}}
  \\
  \parbox{0.85\linewidth}{\SubFig{b}
  \IncFig[clip,width=6.5cm]{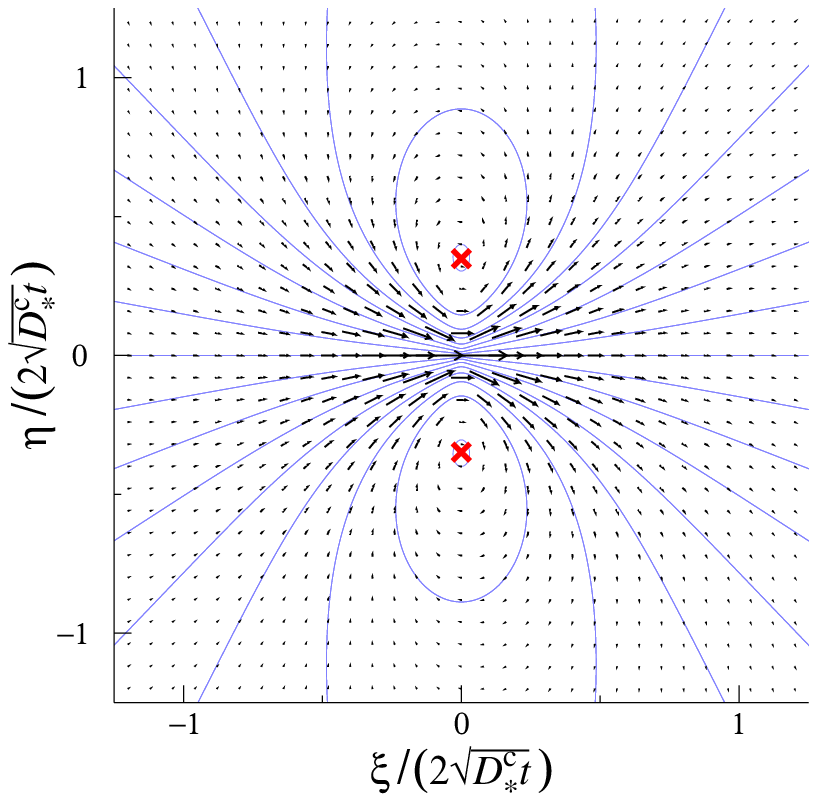}}
  \caption{\label{Fig:RR.mu}%
  Plot of the analytically calculated 2D displacement correlation.
  The value of $\ChiR \cdot \mb{e}_1$
  is plotted as a vector field 
  on the $\BoldXi$-plane 
  (normalized with $2\sqrt{{\Dc_*}t}\,$), 
  together with the ``streamlines''; 
  the crosses indicate the centers of the vortices.
  (a) The result of the linear theory 
  in Eqs.~(\ref{eqs:ChiR.L}).
  (b) A modified plot based on Eqs.~(\ref{eqs:ChiR.mu}), 
  with $\muR = 0.25 > 0$.
  }
\end{figure}

\begin{figure}
  \centering
  \parbox{0.85\linewidth}{\SubFig{a}
  \IncFig{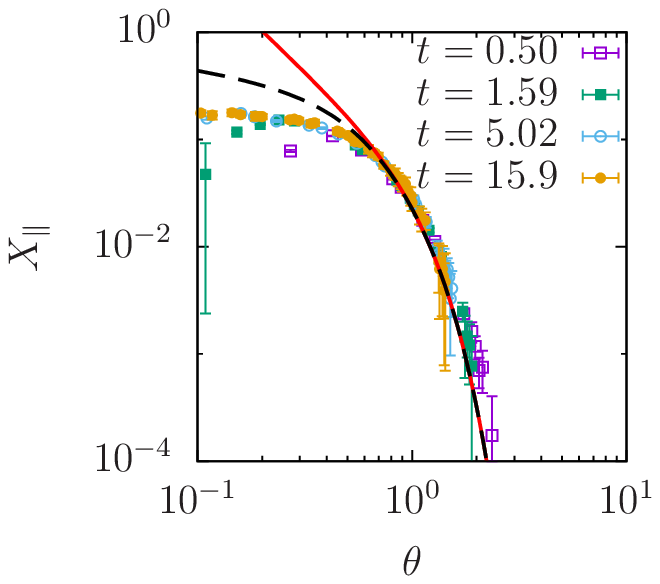}}
  \\
  \parbox{0.85\linewidth}{\SubFig{b}
  \IncFig{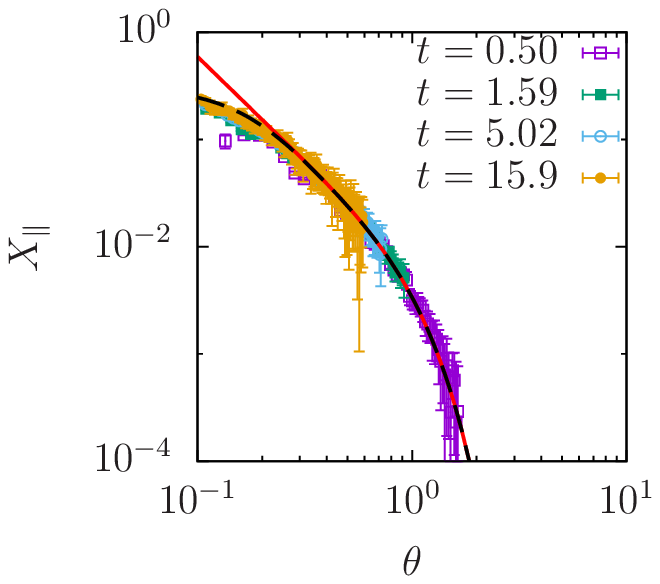}}
  \\
  \parbox{0.85\linewidth}{\SubFig{c}
  \IncFig{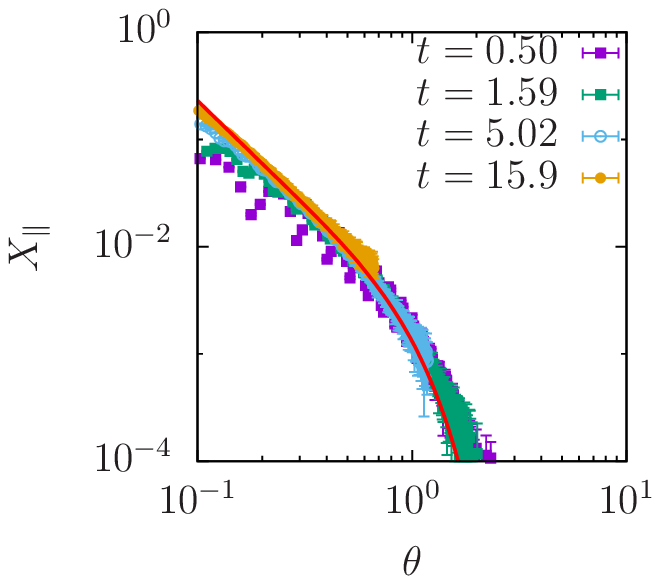}}
  \caption{\label{Fig:phi}%
  Numerical values of $\Xl$ for different values of $\phi$:
  (a) $\phi = 0.4$, (b) $\phi = 0.6$, and (c) $\phi = 0.7$, 
  plotted basically in the same way 
  as in Fig.~\protect\ref{Fig:scale}. 
  In the denser cases (b) and (c), 
  greater values of $\Vmax$ were used,  
  in order to keep the computational condition 
  close to that of hard spheres,
  which was confirmed by checking 
  that the MSD does not depend on $\Vmax$ any more. 
  The values of $\Vmax$ and $S = S(k\to+0)$, 
  as well as that of the fitting parameter $\muR$, 
  are summarized as follows:
  (a) $\Vmax = 50  \,\kT$, $S = 0.246$, $\muR = 0.16$; 
  (b) $\Vmax = 500 \,\kT$, $S = 0.055$, $\muR = 0.052$; and 
  (c) $\Vmax = 5000\,\kT$, $S = 0.025$, $\muR = 0$.
  }
\end{figure}


We have calculated 
  displacement correlation tensor 
  both numerically and analytically, 
  which allows us 
  to capture vortical cooperative motion
  in 2D colloidal liquids.
The analytical calculation predicts the cooperative motion  
  to have a self-similar space--time structure, 
  expressible with the similarity variable $\theta$.
This prediction 
  has been verified numerically 
  and shown to be quantitatively valid 
  for large $\theta$, 
  corresponding to length scales 
  comparable to and greater than $\ld(t)$. 
For shorter scales, 
  the prediction in Eqs.~(\ref{eqs:ChiR.L}) 
  deviates from the computed result, 
  but it remains true 
  that the numerical data sets for different $t$
  collapse onto a single master curve.

As we have stated in the Introduction,  
  these calculations of $\Av{\mb{R}\otimes\mb{R}}$
  are intended to prepare the ground 
  for a more elaborate theory of colloidal liquids.
Granted that the area fraction 
  in the present study ($\phi = 0.50$) 
  is not large enough 
  to exhibit genuine longtime behavior 
  of colloidal glasses, 
  the present results are already indicative 
  of some precursory features of glassy dynamics. 
Let us discuss these features, 
  including the longevity of the vortical motion, 
  negative longtime tail in the velocity autocorrelation, 
  and slowdown of $\Cr$ related to the shear modulus of the liquid.

\subsection{Speculation on caged dynamics}
\label{subsec:cage}

The vortical cooperative motion 
  indicated by $\Av{\mb{R}\otimes\mb{R}}$ 
  is much long-lived, 
  in comparison to the timescale of the structural relaxation 
  (often denoted with $\tau_\alpha$ 
  for glassy systems \cite{Berthier.RMP83,Doliwa.PRE61}),
  measured by the decay of $F(k,t)$
  at the nearest-neighbor distance.
In the present case with $\phi = 0.5$, 
  the decay time of $F(k,t)$ 
  at $k = k_{\text{peak}} = 6.3\,\sigma^{-1}$ %
  is smaller than unity (i.e.\ 
  shorter than $\sigma^2/D$ in dimensional expression), 
  while the correlations persists to grow 
  until $t = 15.9$ at least,
  as is attested by Figs.~\ref{Fig:Chi} and \ref{Fig:scale}.
The mechanism of this (relative) persistence 
  is clarified by Eq.~(\ref{AP2.H}),
  in which the integrals are dominated 
  by the long-wave components 
  whose relaxation time diverges as $k^{-2}$ for $k\to0$.
The persistence of the cooperative motion 
  holds also for denser systems 
  studied by Doliwa and Heuer \cite{Doliwa.PRE61}, 
  as is exemplified by their Fig.~8 
  illustrating the vortical motion 
  at $\phi = 0.77$ and $t = 10\tau_\alpha$. 

The self-similar behavior of $\Av{\mb{R}\otimes\mb{R}}$, 
  expressible with the similarity variable $\theta$, 
  suggests an onion-like or a matryoshka-like structure,
  such that the small cages are nested 
  in larger and slower cages
  \cite{Ooshida.PRE88}.
The time scale $\tau_\alpha$
  signals the start of the cage collapse 
  on the innermost layer of the nested cages, 
  but the outer cage layers survive, 
  though they may be deformed 
  according to the displacement of the central particle.
We note that this structure is also observed  
  in the quasi-1D dynamics of Brownian particles
  (single-file diffusion with overtaking), 
  as is shown in Fig.~4 of Ref.~\cite{Ooshida.MPLB29},
  where the overtaking destroys 
  most of the displacement correlation of neighboring particles
  but there remains very weak positive correlation 
  corresponding to the outer cage layers.



Let us now focus on the central particle
  confined in the cage structure. 
Although the short-range behavior of Eqs.~(\ref{eqs:ChiR.L}) 
  lacks quantitative accuracy, 
  it seems to include 
  an important feature of colloidal systems 
  in regard to the velocity autocorrelation, 
  $\Av{\mb{u}(t)\cdot\mb{u}(s)}$,
  which is shown to be \emph{negative}.
The velocity autocorrelation  
  is a special case 
  of the Lagrangian velocity correlation, 
  $\Av{\mb{u}(\BoldXi,t)\otimes\mb{u}(\BoldXi',s)}$, 
  which is related to the displacement correlation 
  as 
\begin{multline}
  \ds\dt\Av{\mb{R}(\BoldXi,t,s)\otimes\mb{R}(\BoldXi',t,s)} \\{}
  = 
  - \Av{\mb{u}(\BoldXi,t)\otimes\mb{u}(\BoldXi',s)} 
  - \Av{\mb{u}(\BoldXi,s)\otimes\mb{u}(\BoldXi',t)} 
  \label{ddRR.tensor}.
\end{multline}
By taking the trace of Eq.~(\ref{ddRR.tensor}) 
  and considering the limit of $\BoldXi'-\BoldXi \to \mb{0}$, 
  we obtain  
\begin{equation}
  \Av{\mb{u}(t)\cdot\mb{u}(s)}
  = -\frac12\ds\dt
  \Ev{\left[ \Xl(\xi,t,s) + \Xtr(\xi,t,s) \right]}{\xi=0}
  \label{ddRR.2D} 
\end{equation}
  in the 2D case, 
  and substitution of Eqs.~(\ref{eqs:ChiR.L}) 
  yields
\begin{equation}
  \Av{\mb{u}(t)\cdot\mb{u}(0)}
  = -\frac{S}{4\pi\rho_0} t^{-2} < 0 
  \label{uu.2D}.
\end{equation}%
Thus the velocity autocorrelation 
  is shown to have a negative tail, 
  with the exponent ${-2}$ in the present case. 

More generally, 
  in the $\nd$-dimensional case, 
  the expression on right-hand side of the relation 
\begin{equation}
  \Av{\mb{u}(t)\cdot\mb{u}(s)} 
  = -{\frac12} \lim_{\BoldXi \to \mb{0}}
  \ds\dt\Av{\mb{R}(\BoldXi,t,s)\cdot\mb{R}(\mb{0},t,s)} 
  \label{ddRR*}
\end{equation}
  can be evaluated by means of 
  the $\nd$-dimensional AP formula~\cite{Ooshida.BRL11}, 
\begin{widetext}%
\begin{equation}
  \Av{{{R_\alpha}(\BoldXi,t,s)}{{R_\beta}(\mb{0},t,s)}} 
  \propto 
  \int
  \left[ 
  \frac{%
  C_{\alpha\beta}(\mb{k},s,s) + C_{\alpha\beta}(\mb{k},t,t)
  }{2}
  - C_{\alpha\beta}(\mb{k},t,s)
  \right]
  \frac{e^{-\II\mb{k}\cdot\BoldXi}}{{k_\alpha}{k_\beta}}
  {\D^{\nd}\mb{k}}
  \label{AP.nD},
\end{equation}%
\end{widetext}%
  where
  $\alpha,\beta \in \{1,2,\ldots,\nd\}$ and 
\[ 
  C_{\alpha\beta}(\mb{k},t,s) 
  = \frac{\rho_0^2}{N}  
  \Av{{\check\Psi_\alpha(\mb{k},t)}{\check\Psi_\beta(-\mb{k},s)}}
  \quad 
\]
  denotes the correlation of the deformation gradient tensor 
  defined in analogy with Eqs.~(\ref{dr/dXi})--(\ref{2D.m2}).
The correlation $C_{\alpha\beta}$
  in the integrand of Eq.~(\ref{AP.nD}) 
  will comprise exponential terms, 
  analogous to Eqs.~(\ref{d4psi}) and (\ref{Lin2.Cd})
  and making a negative contribution 
  to the velocity autocorrelation, 
  as well as a linear term 
  such as Eq.~(\ref{Lin2.Cr}), whose contribution vanishes. 
Therefore, in total, 
  the velocity autocorrelation 
  must be negative.
The exponent 
  is evaluated as follows: 
Differentiating $C_{\alpha\alpha}$ in Eq.~(\ref{AP.nD}) 
  with $s$ and $t$, 
  under the assumption that the exponential term in it 
  has essentially the same $k$-dependence 
  as Eqs.~(\ref{d4psi}) and (\ref{Lin2.Cd}), 
  we find 
\begin{align}
  -\Av{\mb{u}(t)\cdot\mb{u}(s)} 
  &= {\frac12}\sum_\alpha 
  \dt\ds\Av{{{R_\alpha}(\mb{0},t,s)}{{R_\alpha}(\mb{0},t,s)}} 
  \notag \\%
  &\propto
  \int e^{-\Dc_* {k^2}(t-s)}
  {k^2} \,\D^{\nd}\mb{k}
  \notag \\%
  &\propto 
  (t-s)^{-(\nd+2)/2}
  \label{uu.nD}.
\end{align}
Setting $\nd=2$ in Eq.~(\ref{uu.nD}) 
  reproduces the exponent $-2$ in Eq.~(\ref{uu.2D}).
It is also consistent 
  with the exponent $-5/2$ of the negative longtime tail 
  in the three-dimensional (3D) colloidal systems 
  \cite{Dhont.Book1996},  
  and also with the 1D results 
  \cite{van-Beijeren.PRB28,Taloni.PRE78}
  in which the exponent is $-3/2$.
A result analogous to Eq.~(\ref{uu.nD}) 
  was also reported by Hagen \textit{et al.} \cite{Hagen.PRL78}
  for a fluid confined in the quasi-$\nd$-dimensional space.

The temporal behavior of the velocity autocorrelation 
  in Eqs.~(\ref{uu.2D}) and (\ref{uu.nD})
  clarifies that, 
  in spite of the apparent resemblance 
  of Fig.~\ref{Fig:RR.2D}
  to the back\-flow structure 
  observed in molecular dynamics 
  by Alder and Wainwright \cite{Alder.PRA1}, 
  the two types of vortical cooperative motions 
  originate from totally different mechanisms.
The longtime tail of the velocity autocorrelation 
  in Newtonian--Hamiltonian molecular dynamics
  is basically \emph{positive}, 
  resulting from conservation of vorticity (the rotational 
  part of the momentum field).
Contrastively, 
  the velocity autocorrelation in colloidal systems 
  is seen to have a \emph{negative} tail. 
The negative value of $\Av{\mb{u}(t)\cdot\mb{u}(0)}$
  is a manifestation of the cage effect:
  the particles are always pushed back by its neighbors.
In this sense, 
  the cage effect is successfully captured 
  by the analytically calculated displacement correlations 
  in Eqs.~(\ref{eqs:ChiR.L}), 
  which directly lead to Eq.~(\ref{uu.2D}) 
  describing the negative tail behavior.

Now, recalling our remark in the Introduction 
  that the case of Newtonian molecular dynamics 
  for dense liquids 
  deserves special consideration, 
  we note that the longtime tail in such systems 
  can be rather complicated, 
  as was reported 
  by Williams \textit{et al.}~\cite{Williams.PRL96}.
They studied 
  Newtonian molecular dynamics of 3D hardcore fluids 
  with various densities.
For the lowest density, 
  the velocity autocorrelation 
  remains positive for all $t$, 
  and the Alder--Wainwright longtime tail is clearly observed.  
For liquids with somewhat higher densities,
  the velocity autocorrelation 
  is seen to be transiently negative but becomes positive again.  
For even higher densities in the ``supercooled'' regime, 
  the positive longtime tail vanishes 
  and the negative tail dominates. 
These findings of Williams \textit{et al.}~\cite{Williams.PRL96}
  could be interpreted, in light of Eq.~(\ref{uu.nD}), 
  as follows:
For Newtonian molecular dynamics with moderate densities, 
  asymptotically 
  $\Av{\mb{u}(t)\cdot\mb{u}(0)}$ will be expressible  
  as a superposition of the Alder--Wainwright tail, 
  ${+A} t^{-\nd/2}$, 
  and the negative longtime tail due to the cage effect, 
  ${-B} t^{-(\nd+2)/2}$,  
  so that 
\begin{equation}
  \Av{\mb{u}(t)\cdot\mb{u}(0)} 
  \sim +A t^{-\nd/2} - B t^{-(\nd+2)/2}
  \label{uu.MD}
\end{equation}
  for such systems.
The term with $+A$ is due to the conservation of vorticity,
  while that with $-B$ originates from the mass conservation.
Since $t^{-\nd/2}$ 
  decays more slowly, 
  the first term is expected to survive.  
As the density is increased, 
  the coefficient $A$ decreases and $B$ increases, 
  and finally $A$ disappears (if we accept 
  the conclusion of Williams \textit{et al.}~\cite{Williams.PRL96}
  that the Alder--Wainwright tail was not observed any more).
This interpretation also seems to be consistent 
  with the kinetic theory of monoatomic liquids 
  by Sj{\"o}gren and Sj{\"o}lander \cite{Sjogren.JPhC12},  
  stating that the memory function due to the the density modes 
  and that due to the vorticity modes 
  contribute to the diffusion separately, 
  with the former dominating at higher densities 
  and the latter at lower.


Let us return 
  to the displacement correlation 
  in Eqs.~(\ref{eqs:ChiR.L}).
They  suggest 
  that the MSD has a logarithmic correction term 
  to the normal diffusion~\cite{Ooshida.BRL11}.
If we adopt 
  the estimation 
\begin{equation}
  \Av{\smash{\left[\mb{R}(t)\right]^2}} 
  \sim \Xl(\xi_{\text{cut}},t)
  \label{Xl@cutoff}
\end{equation}
  with the ``cutoff length'' 
  $\ell_0\xi_{\text{cut}}$, 
  Eq.~(\ref{Xl}) yields
\begin{equation}%
  \Av{\mb{R}^2}
  \sim 
  \Da t
  + \frac{\Order(1)}{\rho_0}\ln{\frac{Dt}{\sigma^2}}
  \label{MSD.t+log}
\end{equation}
  with $\Da \sim D$ if $\xi_{\text{cut}} \sim \Order(1)$.

The logarithmic term 
  is characteristic of 2D caged particles 
  and 2D elastic bodies.
The dynamics of caged particles, 
  for timescales shorter than $\tau_\alpha$,
  can be modeled by elastic network 
  \cite{Brito.EPL76,Brito.JCP131};
  according to this modeling 
  and on the assumption that continuum description is valid, 
  the calculation of the MSD 
  reduces to the Edwards--Wilkinson integral 
  \cite{Edwards.PRSLA381,Toninelli.PRE71},
  which leads to the logarithmic behavior.
Recently, 
  another model that relates the caged diffusion 
  to the Edwards--Wilkinson equation 
  was studied by Centres and Bustingorry~\cite{Centres.PRE93}.
Their model 
  consists of diffusing particles on a 2D lattice,
  with a constraint 
  reminiscent of the eight-queens problem 
  \cite{Bell.DiscMath309,Dean.JPA31,Hukushima.CompPhysComm147}.
Since the cages never break in this model, 
  the normal diffusion disappears.
Centres and Bustingorry~\cite{Centres.PRE93}
  showed that the computed MSD,
  in the case 
  corresponding to the 2D version 
  of the unbiased single-file diffusion,  
  is consistent with the prediction 
  of the 2D Edwards--Wilkinson equation.
This behavior, 
  which corresponds to Eq.~(\ref{MSD.t+log}) with $\Da = 0$,
  is in contrast to the more usual 2D lattice dynamics 
  of particles 
  interacting through the excluded volume effect alone
  \cite{Nakazato.PTP64,van-Beijeren.PRL55},
  in which 
  the MSD takes the same form as in Eq.~(\ref{MSD.t+log})
  with finite $\Da$;
  see Eq.~(11) in Ref.~\cite{van-Beijeren.PRL55}.
Besides, in the case of colloidal liquids without lattice,  
  such as the ones 
  modeled by Eqs.~(\ref{Langevin-.r})--(\ref{V=}) 
  with $\nd \ge 2$ and $\Vmax \to +\infty$, 
  Osada \cite{Osada.PrThRF112} 
  has proven that $\Da$ never vanishes 
  for any finite temperature and pressure 
  \cite{Osada.PrThRF112,Berthier.RMP83}.  
In the sense 
  that non-zero $\Da$ is incorporated 
  together with the cage effect, 
  the present analysis seems to be 
  a step in the right direction 
  as a theory of colloidal liquids.

\subsection{Toward mode-coupling theory of elasticity}
\label{subsec:MCT}

Being a kind of four-point space--time correlation, 
  the displacement correlation, $\Av{\mb{R}\otimes\mb{R}}$,
  delivers useful information about dynamics of dense liquids.
The keystone 
  in our analytical calculation of $\Av{\mb{R}\otimes\mb{R}}$
  is the Alexander--Pincus formula:
  it relates the displacement correlation, 
  a \emph{four-point} statistical quantity, 
  to the \emph{two-body} correlation 
  of the Lagrangian deformation gradient tensor.
We have observed 
  that, by way of the 2D Alexander--Pincus formula (\ref{AP2.H}),
  the linear analysis of the Dean--Kawasaki equation
  already gives 
  the longitudinal and transverse displacement correlations 
  in Eq.~(\ref{eqs:ChiR.L}),  
  which captures both the non-zero $\Da$ 
  and the (precursory) caged dynamics.

Nevertheless,
  it is still problematic 
  that the linear theory 
  cannot be accurate enough.
In particular,
  quantitative prediction 
  of $\Da = \lim_{t\to\infty}\Av{\mb{R}^2}/(2{\nd}t)$
  is beyond its scope.
The most serious limitation 
  is found in the short-range behavior of $\Xtr$:
  according to Eq.~(\ref{Xtr}), 
  the value of $\Xtr(\xi,t)$ remains negative 
  for $\xi\to+0$,  
  while it should be positive for sufficiently small $\xi$,
  as is suggested by the numerical results. 
Pictorially speaking,
  the distance 
  between the centers of the vortices 
  in Fig.~\ref{Fig:RR.2D} 
  must be finite, 
  but the linear theory predicts 
  the distance to be zero,
  as is shown in Fig.~\ref{Fig:RR.mu}(a).

As a remedy for this discrepancy, 
  we take notice of Eq.~(\ref{Cr.AP}) 
  describing the contribution of $\Cr$
  to the displacement correlation, 
  and consider modification to it
  based on a nonlinear theory 
  to go beyond Eq.~(\ref{2D-lin.Psi.eqn}).
In the case of hardcore particles, 
  the quadratic nonlinear terms can be calculated concretely, 
  which leads to a set of nonlinear equations
  for $\psiD$ and $\psiR$, 
  in the following form:
\begin{widetext}%
\begin{subequations}%
\begin{align}
  (\dt+\Dc_*\mb{k}^2)\psiD(\mb{k},t)
  &= 
  \sumPQK
  \begin{bmatrix} \psiD^*(\mb{p},t) & \psiR^*(\mb{p},t) \end{bmatrix}
  \begin{bmatrix}
    \V{d}{k}dd{p}{q} & \V{d}{k}dr{p}{q} \\
    \V{d}{k}rd{p}{q} & \V{d}{k}rr{p}{q} 
  \end{bmatrix}
  \TwoVect{\psiD^*(\mb{q},t)}{\psiR^*(\mb{q},t)}
  + \Order(\psi^3)
  + \fd(\mb{k},t)
  \label{dt.psiD},
  \\ 
  \dt\,\psiR(\mb{k},t)
  &= 
  \sumPQK
  \begin{bmatrix} \psiD^*(\mb{p},t) & \psiR^*(\mb{p},t) \end{bmatrix}
  \begin{bmatrix}
    \V{r}{k}dd{p}{q} & \V{r}{k}dr{p}{q} \\
    \V{r}{k}rd{p}{q} & 0 
  \end{bmatrix}
  \TwoVect{\psiD^*(\mb{q},t)}{\psiR^*(\mb{q},t)}
  + \Order(\psi^3)
  + \fr(\mb{k},t)
  \label{dt.psiR},
\end{align}%
\label{eqs:dt.psiH}%
\end{subequations}%
\end{widetext}%
  where $\fd$ and $\fr$ are random forces 
  given by some appropriate linear combinations 
  of $\check{f}_1$ and $\check{f}_2$ 
  in Eq.~(\ref{f.2D}). 
On the basis of a preliminary analysis 
  of Eqs.~(\ref{eqs:dt.psiH}) \cite{Ooshida.JPS15s},
  we may replace Eq.~(\ref{Cr.AP}) 
  with an equation analogous to Eq.~(\ref{Cd.AP}), 
\begin{multline}
  \frac{\Cr(\mb{k},s,s) + \Cr(\mb{k},t,t)}{2} - \Cr(\mb{k},t,s)
  \\ {}
  = \frac{1}{{\muR}L^4}
  \left[ 1 - e^{-\muR{D_*}\mb{k}^2(t-s)} \right]
  \label{Cr.AP.mu}, 
\end{multline}
  introducing a parameter $\muR$.
For $\muR\to0$, 
  Eq.~(\ref{Cr.AP.mu}) reduces to Eq.~(\ref{Cr.AP}). 
Positive values of $\muR$
  changes the behavior 
  for large values of $\Dc_*\mb{k}^2(t-s)$, 
  as Eq.~(\ref{Cr.AP}) diverges linearly 
  while Eq.~(\ref{Cr.AP.mu}) converges to $1/({\muR}L^4)$.
The physical meaning of $\muR$
  is suggested by noticing 
  that Eq.~(\ref{Cr.AP.mu}) 
  has the same structure as Eq.~(\ref{Cd.AP}), 
  with $\muR$ corresponding to $S$ 
  and $D/\muR$ to $\Dc_* = D_*/S$;
  since $S$ is related to the compressibility, 
  introducing $\muR$ in Eq.~(\ref{Cr.AP.mu}) 
  means introducing the shear modulus.
The inverse proportionality of the saturation value of $\Cr$
  to $\muR$ 
  seems to be consistent with this interpretation,  
  as is seen from the recent studies 
  by Klix \textit{et al.} \cite{Klix.PRL109} 
  and Flenner and Szamel \cite{Flenner.PRL114}.
More concretely,
  Eq.~(\ref{Cr.AP.mu}) can be rewritten 
  in terms of $S_4^\perp(\mb{q},t)$ in Eq.~(\ref{S4.Tr}),
  via Eq.~(\ref{inv.AP2.Cr}), 
  as 
\begin{equation}
  S_4^\perp(\mb{q},t-s) 
  \propto   
  \frac{1 - e^{-\muR{D}{q^2}(t-s)}}{{\muR}q^2}
  \label{S4.Tr.mu};
\end{equation}
  assuming Eq.~(\ref{Cr.AP.mu}) therefore implies 
  that,
  for $\ell_0^2 \ll q^{-2} \ll {\muR}D(t-s)$, 
  asymptotically $S_4^\perp(\mb{q},t-s)$ 
  should 
  behave in inverse proportion to ${\muR}q^2$.
Indeed, 
  this behavior of $S_4^\perp(\mb{q},t-s)$ 
  was recently demonstrated 
  by Flenner and Szamel \cite{Flenner.PRL114},
  both for glassy solids and liquids.
In the latter case (dense liquids), however,
  it should be noted 
  that the $q^{-2}$ behavior was observed only transiently,
  and therefore the validity of Eq.~(\ref{Cr.AP.mu}) 
  should be limited to some long but finite range of time.

Now let us study  
  how the non-zero $\muR$ modifies 
  the analytical expression of the displacement correlation.   
By substituting Eq.~(\ref{Cr.AP.mu})
  into the Alexander--Pincus formula (\ref{AP2.H}), 
  together with Eq.~(\ref{Cd.AP}), 
  Eqs.~(\ref{eqs:ChiR.L}) are modified as
\begin{widetext}%
\begin{subequations}
\begin{align}
  \Xl(\xi,t) 
  &= 
  \frac{S}{{4\pi}\rho_0}
  \left[
  E_1({\theta^2}) + E_1(\theta^2/\muR)
  + \frac{e^{-\theta^2} - e^{-\theta^2/\muR}}{\theta^2}
  \right]
  \label{Xl.mu},  
  \\
  \Xtr(\xi,t) 
  &= 
  \frac{S}{{4\pi}\rho_0}
  \left[
  E_1({\theta^2}) + E_1(\theta^2/\muR)
  - \frac{e^{-\theta^2} - e^{-\theta^2/\muR}}{\theta^2}
  \right]
  \label{Xtr.mu}
  \relax.
\end{align}%
\label{eqs:ChiR.mu}%
\end{subequations}%
\end{widetext}%
Equation (\ref{Xtr.mu}) has the desirable property 
  that $\Xtr$ is positive for small $\theta$,
  so that the two vortices 
  are now separated by a finite distance,
  as is shown in Fig.~\ref{Fig:RR.mu}(b).
In addition, 
  $\Xl$ in Eq.~(\ref{Xl.mu}) also improves upon Eq.~(\ref{Xl}),  
  as is indicated 
  by the broken line in Fig.~\ref{Fig:scale},
  where $\muR = 0.13$
  gives the best agreement.

Although $\muR$ is treated here as a fitting parameter,
  it should be determined from a nonlinear theory for $\Psi$, 
  namely the mode-coupling theory
  for $\psiD$ and $\psiR$. 
A nonlinear theory 
  is needed also because
  the asymptotic behavior of Eq.~(\ref{Xl.mu}) 
  for $0 < \theta^2 \ll \muR < 1$ 
  implies logarithmic MSD with $\Da = 0$, 
  which is the behavior of a soft elastic body 
  but not that of a colloidal liquid.
While Eq.~(\ref{Cr.AP.mu}) converges
  and Eq.~(\ref{Cr.AP}) diverges 
  rapidly for large $\Dc_*\mb{k}^2(t-s)$,
  the truth is probably somewhere in between,
  as was suggested 
  by the numerical behavior of $S_4^\perp(\mb{q},t)$
  computed for glassy liquids \cite{Flenner.PRL114}.
We expect that a nonlinear theory for $\Psi$ 
  will reveal the correct behavior of $\Cr$, 
  possibly predicting slow divergence
  of the expression on the left-hand side 
  of Eqs.~(\ref{Cr.AP}) and (\ref{Cr.AP.mu}). 

The idea of MCT for displacement correlation
  may sound infeasible, but in fact, 
  the way is already paved to a certain extent.
Firstly, 
  there is no need to develop a closure for four-body correlations, 
  because \emph{two-body} correlations, $\Cd$ and $\Cr$, 
  are sufficient 
  as input into the Alexander--Pincus formula (\ref{AP2.H});
  this is the advantage of the Lagrangian description
  \cite{Ooshida.PRE88,Ooshida.BRL11}.
Secondly,  
  the Lagrangian description 
  has another advantage 
  of changing the multiplicative noise 
  in the Dean--Kawasaki equation into an additive one, 
  which facilitates field-theoretical formalisms of MCT. 
Thirdly, 
  the coefficients in Eqs.~(\ref{eqs:dt.psiH}) 
  can be concretely expressed, 
  at least in the case of 2D hardcore particles. 
Encouraged by these conditions,  
  and also motivated by the intuition 
  that a theory about correlations of deformation gradient tensor 
  will open a door to the rheology of glassy liquids, 
  we have started 
  a preliminary analysis of an MCT-like equations 
  for $\Cd$ and $\Cr$ \cite{Ooshida.JPS15s}.  
An effort in this direction is now in progress 
  and will be reported elsewhere.


Finally, 
  in order to see 
  how the validity of the present theory 
  depends on the area fraction, 
  we computed $\Xl$ for three different values of $\phi$, 
  in addition to $\phi = 0.5$ reported in Sec.~\ref{sec:compare}.
The numerical results 
  are summarized in Fig.~\ref{Fig:phi}, 
  where $\Xl$ is plotted against the similarity variable $\theta$.
It seems to be valid in all the cases---or, 
  at least, except for the case of $\phi=0.7$ 
  in which evidence may be insufficient---%
  that $\Xl(\xi,t)$ for different $t$ is expressible 
  through the similarity variable $\theta$,  
  and reasonable agreement is found near $\theta\sim 1$.
Although the discrepancy 
  in the short-range behavior   
  remains in all the cases, 
  the remedy with $\muR$ seems to be effective 
  for $\phi = 0.6$ 
  [the broken line in Fig.~\ref{Fig:phi}(b)].
In the high-density case ($\phi = 0.7$), 
  while the short-range discrepancy  
  is small for $t = 15.9$,
  the correlation for $\theta > 1$ 
  becomes somewhat greater 
  than the prediction 
  of the linear theory in Eq.~(\ref{Xl.far});
  the behavior seems to approach the exponential decay,
  $\Xl\propto e^{-\theta}$, 
  in agreement with Doliwa and Heuer \cite{Doliwa.PRE61}. 
This enhancement of the displacement correlation  
  for $\theta > 1$ in the high-density case  
  is out of the scope of the linear theory, 
  but the self-similarity still seems to hold, 
  raising the hope 
  that the analytical method in Sec.~\ref{sec:theory} 
  could be somehow extended to this case,
  possibly in the form of field-theoretical development
  of MCT-like equations
  and their approximate solution in terms of a similarity variable.


\section{Concluding remarks}
\label{sec:conc}

Starting from the Dean--Kawasaki equation (\ref{DK.Q})
  that gives a mesoscopic continuum description 
  of colloidal liquids, 
  we have developed analytical calculation 
  of the displacement correlation tensor 
  $\Av{\mb{R}\otimes\mb{R}}$,  
  and compared the analytical results 
  with direct numerical simulation 
  of interacting Brownian particles.
The results of the linear analysis, 
  given in Eqs.~(\ref{eqs:ChiR.L}),  
  already include information 
  about the basic space--time structure of the cooperative motion,
  capturing its self-similar character, 
  the vortical flow pattern, and the negative longtime tail.

The short-range behavior, however, 
  needs to be improved 
  on the basis of a nonlinear analysis.
While the full analysis of Eqs.~(\ref{eqs:dt.psiH})
  clearly belongs to our future work, 
  for the present we have shown a numerical evidence 
  that the analytical results are remarkably improved 
  by introducing $\muR$ via Eq.~(\ref{Cr.AP.mu}).
This is equivalent 
  to introducing some finite value of shear modulus.
As a result, 
  Eqs.~(\ref{eqs:ChiR.L}) 
  are replaced with Eqs.~(\ref{eqs:ChiR.mu}), 
  and the centers of the two vortices, 
  shown in Fig.~\ref{Fig:RR.mu}(b), 
  are now separated by a finite distance, 
  in agreement with the numerical plot 
  in Fig.~\ref{Fig:RR.2D}. 

The present approach 
  has several advantages.
Among various four-point space--time correlations, 
  the displacement correlation $\Av{\mb{R}\otimes\mb{R}}$
  seems to be the most intuitive one.  
The calculated results 
  allow pictorial presentation 
  as in Figs.~\ref{Fig:RR.2D} and \ref{Fig:RR.mu}, 
  which must be helpful 
  in intuitive understanding of cooperative motions.
In previous studies of colloidal systems, 
  this comprehensibility of the displacement correlation 
  was available only at a great cost,    
  as it needed to be calculated from particle-based data.
Now we have discovered a route 
  connecting the displacement correlation tensor 
  and the Dean--Kawasaki equation,  
  which, in principle, makes it possible 
  to calculate $\Av{\mb{R}\otimes\mb{R}}$ 
  without resorting to direct numerical simulation of particles.

From the viewpoint of the future work,   
  in which the displacement correlation will be calculated 
  on the basis of the nonlinear equations (\ref{eqs:dt.psiH}), 
  the main significance of the present article 
  would be its methodology. 
It has laid foundation for tensorial MCT 
  to be developed in this future work.
The adoption of the Lagrangian description 
  has made it possible to calculate $\Av{\mb{R}\otimes\mb{R}}$,
  which itself is a four-point correlation,
  from two-body correlations such as $\Cd$ and $\Cr$. 
The Lagrangian description 
  has also expelled the $\rho$-dependence 
  from the random force of the Dean--Kawasaki equation, 
  thus removing an obstacle 
  for the field-theoretical development of MCT.

We must admit, on one hand,
  that still some issues are to be settled 
  before the plan of tensorial MCT is realized.
Due to the change of variables,
  the rewritten version of the Dean--Kawasaki equation 
  has an infinite number of nonlinear terms.
It is not clear under what condition 
  the higher-order terms can be ignored.
Probably some of the terms must be retained 
  so as to respect certain kinds of symmetry, 
  such as those with regard to time-translation 
  and relabeling. 
It is also necessary 
  to develop equations that determine 
  the ``initial values'' of $\Cd(\mb{k},t,s)$ and $\Cr(\mb{k},t,s)$
  for $t=s$, 
  because it is not evident 
  how the initial value of $\Cd$ is related to $S$ in general, 
  and also because $\Cr$ 
  seems to involve a kind of apparent age\-ing.

On the other hand, 
  if these issues are settled successfully 
  or turn out to be harmless, 
  the tensorial MCT will be quite fruitful.
It will shed light of analytical treatments   
  on the cooperative dynamics in glassy liquids.
By determining the behavior of $\Cr$  
  and thus giving the value of $\muR$,  
  it provides information of the shear modulus, 
  and its extension with shear flow 
  will give a direct access 
  to the rheology of colloidal suspensions.
Besides, since Eqs.~(\ref{eqs:dt.psiH}) have nonlinear terms 
  that couple $\psiD$ and $\psiR$ mutually, 
  the vortical motion involving $\Cr$ 
  may influence the dynamics of $\Cd$ that govern the cage collapse.
The tensorial MCT may clarify 
  how this coupling modifies the predictions of the standard MCT 
  based on the dilatational modes alone. 
We hope that the present work  
  will contribute to 
  this quite intriguing nonlinear theory, 
  in which both the dilatational and the rotational modes 
  are coupled with each other 
  under the over\-damped Dean--Kawasaki dynamics.


\begin{acknowledgments}
  We express our cordial gratitude 
  to Hajime Yoshino, Atsushi Ikeda, 
  Takahiro Hatano, Kunimasa Miyazaki, 
  Ferenc Kun, Grzegorz Szamel, 
  Tadashi Muranaka, Hisao Hayakawa, Ken Sekimoto, 
  Akio Nakahara and So Kitsunezaki 
  for fruitful discussions, 
  including thought-provoking questions and insightful comments. 
  We also thank Alexander Mikhailov,
  not only for such discussions, 
  but also 
  for pointing out some old instances 
  of diffusion equation with multiplicative noise, 
  including Ref.~\cite{Gardiner.Book2009} 
  whose first edition dates back to 1983.
  Anonymous referees are acknowledged 
  for helpful comments 
  and for drawing our attention to Ref.~\cite{Flenner.PRL114}.
  This work was supported 
  by Grants-in-Aid for Scientific Research (\textsc{Kakenhi})  
  (C) No.~24540404 and No.~15K05213, JSPS (Japan).
\end{acknowledgments}



\providecommand{\newblock}{\relax}
\bibliography{diffusion,sgm,turbulence,granular,statmech,Book,%
              ref1603,ref1607,note1603-,note1606-}

\begin{thebibliography}{89}
\bibitem[{not({\natexlab{a}})}]{note.V}
\bibinfo{note}{For the interaction potential $V_{jk}$ we could have used any
  softcore potential that has a proper hardcore limit. Here we adopted the
  repulsive harmonic sphere in Eq.~(\ref{V=}), simply because its hard\-core
  limit has been investigated systematically~\cite{Berthier.EPL86}.}
\bibitem[{not({\natexlab{c}})}]{note.number}
\bibinfo{note}{If $\Vmax$ is finite and allows overtaking, it is necessary to
  specify the timing of renumbering to make $\Av{{R_i}{R_j}}$ meaningful
  \cite{Ooshida.MPLB29}.}
\bibitem[{not({\natexlab{d}})}]{note.chain}
\bibinfo{note}{Once the relation between $x$ and $\xi$ is established, the
  problem of rewriting (fluctuating) hydrodynamic equations such as
  Eqs.~(\ref{+cont}) and (\ref{DK.Q}) into the form of Eq.~(\ref{cont.psi}) is
  simply an exercise in change of variables. See the procedure following
  Eq.~(3.7) in Ref.~\cite{Ooshida.JPSJ80}, as well as Appendix 5 in
  Ref.~\cite{Spohn.JStatPhys154}.}
\bibitem[{not({\natexlab{e}})}]{note.C0}
\bibinfo{note}{In the calculation leading to Eq.~(\ref{R1*R2.NL}), the value of
  $C(k,0)$ is needed as the initial value of $C(k,t)$. We can assume $C(k,0) =
  S(0)/L^2$, which seems to be valid within the longwave approximation.}
\end{thebibliography}


\end{document}